\documentclass{article}

\usepackage{arxiv}

\usepackage[utf8]{inputenc} 
\usepackage[T1]{fontenc}    
\usepackage{url}            
\usepackage{booktabs}       
\usepackage{amsfonts}       
\usepackage{nicefrac}       
\usepackage{microtype}      
\usepackage{amsmath} 
\usepackage{graphicx}
\usepackage{doi}
\usepackage{natbib}
\usepackage{svg} %
\usepackage{adjustbox} %
\usepackage{multirow}%
\usepackage{fontawesome}%
\usepackage{pdflscape} 
\usepackage{lscape} %
\usepackage{rotating}
\usepackage{array}%
\usepackage{algorithm}%
\usepackage{algpseudocode}%
\usepackage{listings}%
\usepackage{algorithmicx}%
\usepackage{threeparttable}%

\usepackage{float}         
\usepackage{caption}       
\usepackage{subcaption} %
\usepackage{tabularx}%
\usepackage{graphicx}      
\usepackage{array}         
\usepackage{fontawesome}   
\newcolumntype{P}[1]{>{\centering\arraybackslash}p{#1}}
\title{CDE-Mapper: Using Retrieval-Augmented Language Models for Linking Clinical Data Elements to Controlled Vocabularies}

\author{ \href{https://orcid.org/0000-0001-7292-1026}{\includegraphics[scale=0.06]{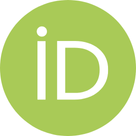}\hspace{1mm}Komal Gilani}\thanks{Use footnote for providing further
		information about author (webpage, alternative
		address)---\emph{not} for acknowledging funding agencies.} \\
	Institute of Data Science\\
	Maastricht University\\
	Maastricht, Netherlands \\
	\texttt{komal.gilani@maastrichtuniversity.nl} \\
	\And
	\href{https://orcid.org/0000-0002-3361-3722}{\includegraphics[scale=0.06]{orcid.png}\hspace{1mm}Marlo Verket} \\
	Department of Internal Medicine I\\
	University Hospital RWTH Aachen\\
	Aachen, Germany  \\
	\texttt{mverket@ukaachen.de} \\
    \And
	\href{https://orcid.org/0009-0007-0263-1470}{\includegraphics[scale=0.06]{orcid.png}\hspace{1mm}Christof Peters} \\
	Department of Cardiology\\
	Maastricht University Medical Center\\
	Maastricht, Netherlands  \\
	\texttt{christof.peters@mumc.nl} \\
    \And
	\href{https://orcid.org/0000-0003-4727-9435}{\includegraphics[scale=0.06]{orcid.png}\hspace{1mm}Michel Dumontier} \\
	Institute of Data Science\\
	Maastricht University\\
	Maastricht, Netherlands \\
	\texttt{michel.dumontier@maastrichtuniversity.nl} \\
     \And
	\href{https://orcid.org/0000-0002-4356-8566}{\includegraphics[scale=0.06]{orcid.png}\hspace{1mm}Hans-Peter Brunner-La Rocca} \\
	Department of Cardiology\\
	Maastricht University Medical Center\\
	Maastricht, Netherlands  \\
	\texttt{hp.brunnerlarocca@mumc.nl} \\
    \And
	\href{https://orcid.org/0000-0003-2817-3950}{\includegraphics[scale=0.06]{orcid.png}\hspace{1mm}Visara Urovi} \\
	Institute of Data Science\\
	Maastricht University\\
	Maastricht, Netherlands \\
	\texttt{v.urovi@maastrichtuniversity.nl} \\
}

\renewcommand{\shorttitle}{\textit{arXiv} Template}

\hypersetup{
pdftitle={A template for the arxiv style},
pdfsubject={q-bio.NC, q-bio.QM},
pdfauthor={David S.~Hippocampus, Elias D.~Striatum},
pdfkeywords={First keyword, Second keyword, More},
}

\begin{document}
\maketitle

\begin{abstract}
	The standardization of clinical data elements (CDEs) aims to ensure consistent and comprehensive patient information across various healthcare systems. Existing methods often falter when standardizing CDEs of varying representation and complex structure, impeding data integration and interoperability in clinical research. We introduce CDE-Mapper, an innovative framework that leverages Retrieval-Augmented Generation approach combined with Large Language Models to automate the linking of CDEs to controlled vocabularies. Our modular approach features query decomposition to manage varying levels of CDEs complexity, integrates expert-defined rules within prompt engineering, and employs in-context learning alongside multiple retriever components to resolve terminological ambiguities. In addition, we propose a knowledge reservoir validated by a human-in-loop approach, achieving accurate concept linking for future applications while minimizing computational costs. For four diverse datasets, CDE-Mapper achieved an average of 7.2\% higher accuracy improvement compared to baseline methods. This work highlights the potential of advanced language models in improving data harmonization and significantly advancing capabilities in clinical decision support systems and research.
\end{abstract}

\keywords{Common data elements \and Clinical data elements \and In-context learning \and Retrieval-Augmented Generation \and Data standardization}

\section{Introduction}

In the digital era of healthcare care, the strategic use of data from electronic health records (EHR), clinical trials, and observational studies is pertinent to advance clinical research and improve healthcare outcomes \cite{kalankesh2024utilization}. However, the potential of these data is frequently not realized due to significant challenges in data standardization and integration across diverse healthcare systems \cite{Hutchings2020A}. An important issue is inconsistent representation and interpretation of clinical concepts across datasets, which requires the use of concept linking. Concept linking is the process of aligning clinical data elements in controlled vocabularies to achieve semantic consistency across systems. This task is also termed entity linking and entity normalization of clinical/biomedical labels. 

Several studies have highlighted the risks associated with poor concept linking. For example, Abigail et al. \cite{whitlock2024icd} identified significant discrepancies between patient cohorts defined by ICD-10 codes and those confirmed by laboratory results, while Hardy et al. \cite{Hardy2022DataConsistency} reported substantial inconsistencies in the recording of ICD-10 codes in conditions such as autism, diabetes, and Parkinson’s disease. These findings highlight the broader challenge of inconsistent concept linking practices, which impede accurate data integration, reproducibility in research, and reliable healthcare decision making. Although no system can entirely eliminate such discrepancies, improved concept linking methods can reduce inconsistencies and enhance interoperability across datasets.

Clinical data elements (CDEs), whether atomic (representing a single characteristic) or composite (capturing multiple interdependent attributes), are foundational to clinical data documentation. Although atomic CDEs can be mapped to established terminologies, composite CDEs introduce additional complexities due to the relationships among attributes (e.g. biomarkers with different measurement methods over time, history of disease pertaining to a family member, etc.). The composite representation of CDEs is often not considered adequately in the concept linking task \cite{entity_linking_overview,Shen2021Entity,Guven2023Multilingual,tabulardataEL,SNOBERT}. A systematic and rigorous approach is required to harmonize these data elements in heterogeneous datasets, ultimately enabling accurate interpretation, transparency, integration, and reuse across diverse healthcare systems and research studies. 

Many studies have explored methods to address the prevalent issue of CDEs standardization in the healthcare domain. Existing approaches range from recurrent neural networks to more advanced deep neural models \cite{ML_Bio_NLP,Loureiro2020MedLinker}, including fine-tuned language models such as bidirectional encoder representations of transformers (BERT) \cite{Perera2020Named}, which can analyze both free text and tabular data \cite{SNOBERT,tabulardataEL}. In particular, ensemble methods \cite{retrieval_based_diagnostic_support}, further refined through fine-tuning for free text (e.g., clinical notes), have been investigated. Traditional rule-based systems such as QuickUMLS \cite{quickumls}, MetaMapLite \cite{MetaMapLite} and cTAKES \cite{CTAKES} have been proposed. Despite these advances, many existing approaches are optimized for atomic CDEs and, therefore, are less effective in standardizing composite CDEs. Additionally, the scalability of these methods is challenged by the need to process large controlled vocabularies, as well as handling overlapping and granular concepts.

Large language models (LLMs), also known as generative models, are a class of artificial intelligence models designed to understand and generate human-like text based on vast amounts of training data \cite{zhao2023survey,naveed2023comprehensive}. LLMs can capture complex language patterns and contextual relationships in data from various modalities \cite{zhao2023survey,thirunavukarasu2023large}. They have been widely used in natural language processing tasks such as text generation, translation, summarization, and question answering, demonstrating their versatility and effectiveness in handling diverse linguistic challenges \cite{zhao2023survey,thirunavukarasu2023large,chang2024survey}. However, LLMs generally lack intrinsic domain knowledge unless they are fine-tuned or augmented with domain-specific resources \cite{Lewis2020Retrieval-Augmented}. This limitation becomes apparent in clinical data standardization, where specialized terminologies, varying data representations, and large, evolving controlled vocabularies pose significant challenges \cite{chang2024survey}.

To address these challenges, existing research has explored zero-shot learning \cite{thapa2023chatgpt} and schema-guided prompts \cite{,SPIRES} to help LLM align extracted information with structured knowledge bases, even without task-specific training data. In particular, Retrieval-Augmented Generation (RAG) has emerged as a promising technique \cite{Zhang2023Siren's,Lewis2020Retrieval-Augmented}, as it combines the generative capabilities of LLM with non-parametric memory (an external resource) to dynamically retrieve relevant domain-specific knowledge \cite{gao2024retrievalaugmentedgenerationlargelanguage}. By accessing large, frequently updated controlled vocabularies on demand, RAG ensures that concept linking remains accurate, up-to-date, and better aligned with the multifaceted requirements of clinical data standardization. This ability to seamlessly integrate external knowledge makes RAG particularly well suited for handling complex or evolving terminologies, which is central to our approach.

In this work, we introduce a novel RAG model that tackles two major challenges: (1) inconsistent representation of terminologies across data dictionaries and (2) the inherent complexity of composite CDEs. Our approach surpasses existing baselines by comprehensively standardizing both atomic and composite CDEs, addressing key limitations not fully resolved in prior studies \cite{liu2021selfalignment,promptlink_MCN,kriss_bert}. Specifically, we handle varying levels of granularity and dependency in CDEs, overlapping terminologies, and the integration of large-scale knowledge bases for enhanced concept linking.
To achieve this, we develop a modular RAG architecture comprising query decomposition, ensemble retrieval, knowledge filtering, and a knowledge reservoir. Critically, our knowledge reservoir incorporates a validation mechanism to verify mapped concepts before storing them for future use, ensuring scalability for real-time applications. This validated modular design distinguishes our approach from existing solutions by offering a more robust framework for accurate and efficient standardization of CDEs.

The remainder of this paper is organized as follows. Section \ref{2} describes the background and related work, Section \ref{sec3} presents our methodology and proposed framework, Section \ref{4} and \ref{sec5} detail the experimental setup and results, Section \ref{sec6} provides interpretation of results and discusses the limitations, and Section \ref{sec7} concludes the paper.

\section{Background and Related Work}\label{2}
\subsection{Clinical Data Elements (CDEs)}

\begin{figure}[ht]
\centering
\includegraphics[width=1\textwidth]{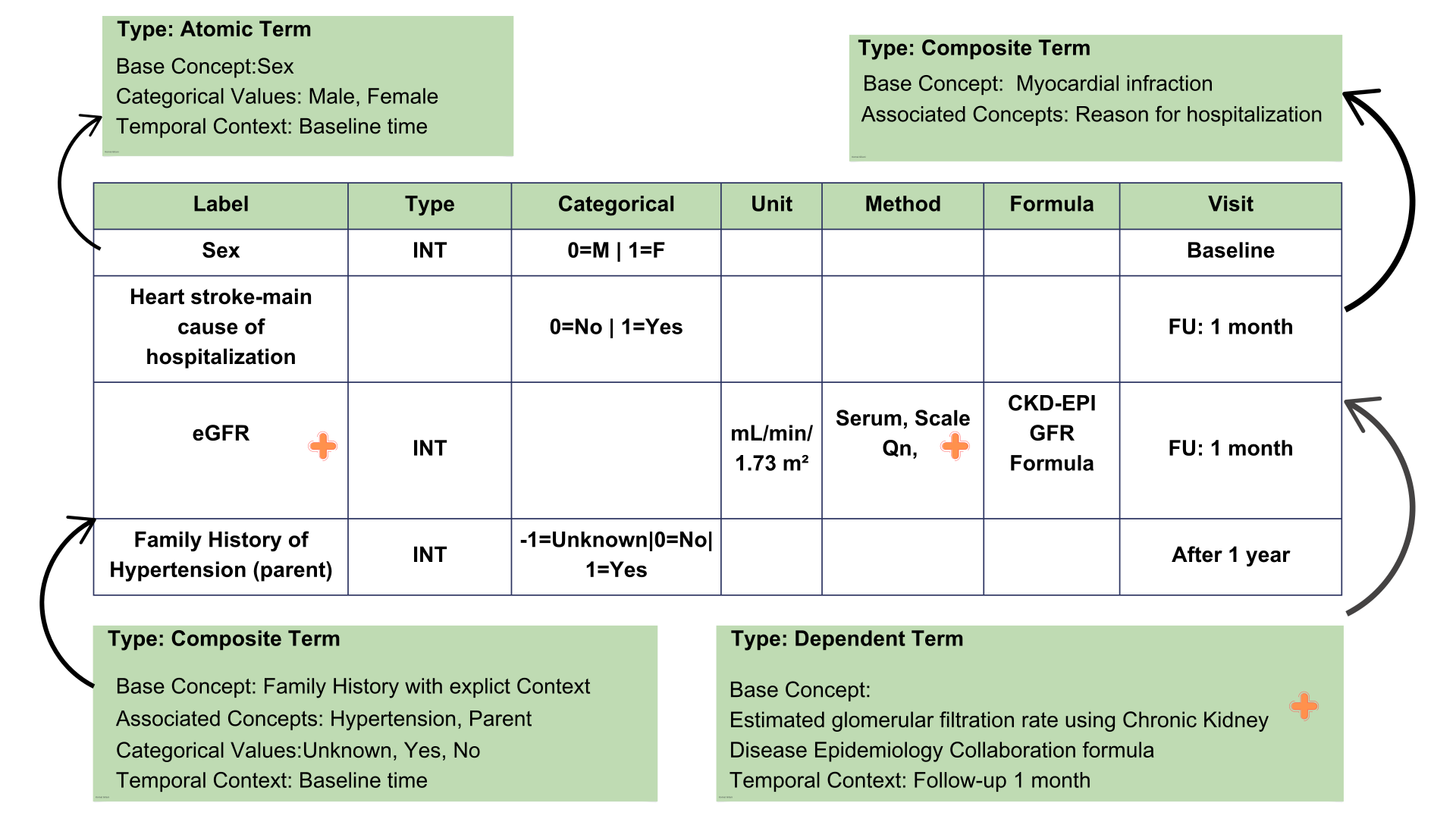}
\caption{Atomic and Composite terms for for 4 example clinical data elements (sex, GGT, FHH, Caisson). The CDEs are described by their common label, the encoding datatype (e..g int), the nature (e.g. discrete or continuous), associated unit, point in time, method, or formula.}\label{fig1}
\end{figure}

Clinical Data Elements are the fundamental building blocks of healthcare information, describing patient demographics, diagnoses, laboratory tests, and more \cite{sheehan2016improving}. These elements reside in data dictionaries that provide metadata such as data types, permissible values, associated attributes, and definitions. Standardization of CDEs involves linking them to controlled vocabularies (e.g., ICD-10, SNOMED CT), ensuring consistent interpretation and interoperability between disparate systems \cite{le2020challenges}. 
As illustrated in Fig.~\ref{fig1}, CDEs vary in their representation and structure:
\begin{itemize}
 \item Atomic CDEs represent a single characteristic (for example, 'sex' or 'blood group'). Even seemingly simple CDEs can be encoded differently (e.g. M / F vs. 0/1), leading to inconsistencies if not harmonized.
 \item Composite CDEs include interdependent or hierarchical attributes, for example, a family history entry that documents both the specific diagnosis and details about the affected family member, etc. A pertinent example of dependent CDEs is biomarkers, which are measured at specific intervals using different techniques. These composite structures add layers of complexity because multiple attributes must be accurately assigned to standardized terminologies \cite{compositeCDEs}.
\end{itemize}
Accurate representation of these CDEs, especially composites, is vital for downstream tasks such as data integration, clinical analytics, and AI-based clinical decision support.

\subsection{Controlled Vocabularies}
One effective approach to ensure consistent representation of CDEs is the adoption of Common Data Models (CDM) and their associated controlled vocabularies. CDMs aim to mitigate the challenges of heterogeneity in healthcare systems by defining standardized data structures and terminologies. The most prominent open-source CDMs, such as FDA Sentinel \cite{platt2018fda} and OMOP \cite{klann2019data}, vary in their support of structural and semantic standardization. Among these, the OMOP Common Data Model \cite{OMOP_advancement}, developed by the Observational Health Data Sciences and Informatics (OHDSI) community, is increasingly used to structure and standardize observational health data in various clinical and research settings \cite{interoperability_in_digital_medicine}. 

In OMOP, data are organized into a person-centric relational database format that captures a wide range of clinical domains, including medical conditions, procedures, medications, healthcare visits, events, devices, demographics, and family history \cite{OMOP_advancement,klann2019data}. Central to OMOP’s adoption is its extensive use of controlled vocabularies, which maintain terminological consistency across studies. By mapping CDEs to these vocabularies, researchers and clinicians can align different data sources more easily. This alignment facilitates transparent reporting and enhances the reproducibility of predictive models.

In this work, we have leveraged a knowledge base of more than three million concepts from controlled vocabularies such as SNOMED, LOINC, ICD, ATC, MedDRA, RxNorm/RxNorm-Extension, MeSH, UCUM, and OMOP Extension and Genomic. These vocabularies collectively cover a wide range of domains: conditions, procedures, medications, demographics, measurements, etc., thus supporting the harmonization of complex CDEs across multiple healthcare systems \cite{design_implemented_standardized_framework_patient_level_prediction}. The widespread adoption of these controlled vocabularies also improves the reproducibility of our work and facilitates data sharing within the broader research community, which is crucial to advance medical knowledge and predictive modeling \cite{omop_EHR_phenotyping,omop_rare_diseases,omop_standardizing_pulmonary_hypertension}.

\subsection{Retrieval-Augmented Generation (RAG)}
For effective linking of CDEs in controlled vocabularies, there is still a pressing need for dynamic and context-aware methods. Retrieval-augmented generation (RAG) models meet this need by combining pre-trained parametric knowledge with dynamically retrieved non-parametric information \cite{Lewis2020Retrieval-Augmented,few_shot_RAG}. It enables language models to generate more accurate and contextually relevant responses \cite{Lewis2020Retrieval-Augmented,Zhang2023Siren's}. This approach is particularly effective for addressing LLM limitations, such as hallucinations and difficulty accessing specific background knowledge, especially when working with specialized domains such as clinical data \cite{Zhang2023Siren's,Lewis2020Retrieval-Augmented,gao2024retrievalaugmentedgenerationlargelanguage}.

In-context learning, a technique that allows models to adapt to new tasks by leveraging examples embedded directly in the input prompt, further enhances the capabilities of RAG systems \cite{rubin2021learning}. The efficacy of in-context learning depends on the quality of the training examples provided; specifically, the relevance and diversity of these examples are critical factors that determine the performance of the model during inference \cite{survey_in_context_learning,ICL_survey}. Recent methods have focused on optimizing the selection of examples to maximize the effectiveness of in-context learning. Relevance-based approaches \cite{incontext_examples_gpt3} prioritize examples aligned with the input query, while techniques like k-nearest neighbor prompting \cite{knn_prompting} identify semantically similar examples from a predefined set. Similarly, advancements in the RAG architecture include a modular approach that allows flexible customization and optimization of the retrieval and generation components \cite{Lewis2020Retrieval-Augmented,gao2024retrievalaugmentedgenerationlargelanguage}. Other approaches prioritize dense retrieval using semantic embeddings or sparse retrieval using keyword-based matching or graph-based methods \cite{benchamrking_RAG}. 

\subsection{Embedding Models}
Transformer-based embedding models \cite{wolf2019huggingface} create rich semantic spaces that are crucial to retrieve relevant information in RAG systems. By converting text into high-dimensional vectors that capture latent semantic features, these models leverage pre-trained knowledge—further refined through fine-tuning—to anchor and interlink biomedical concepts. In this vector space, semantically related terms cluster together, facilitating efficient retrieval based on spatial proximity \cite{semantic_Similarity_evoluation}.

However, models trained on general corpora often fall short in accurately representing specialized clinical/biomedical terms. To address this gap, self-alignment pretraining techniques like SapBERT \cite{liu2021selfalignment} tailor embedding spaces specifically for clinical/biomedical entities. By minimizing contrastive loss, SapBERT draws semantically similar entities closer, thus improving precision in identifying relevant medical information. Sparse embedding models such as SPLADE \cite{spladev2} offer an additional layer of keyword-level matching, complementing dense embeddings in biomedical retrieval tasks where string match suffice for concept linking. Within RAG systems, these refined embeddings enable more effective retrieval by mapping queries to the most contextually relevant medical concepts in controlled vocabularies, thus enhancing both accuracy and specificity in concept linking.

\subsection{Related Work}
One of the common approaches to concept linking is representation via lookup tables such as the UMLS browser and the rule-based system that can capture semantic lexicons to identify concepts in text such as cTAKES \cite{CTAKES}, MetaMap \cite{MetaMapLite} etc. Recent advances in machine learning, deep learning \cite{ML_Bio_NLP}, and LLM have significantly improved the performance of concept linking. In recent years, many deep learning-based methods have been proposed, including single neural network-based, multitask learning-based, transfer learning-based, and hybrid model-based methods. These methods have shown state-of-the-art performance in identifying clinical/biomedical concepts from various datasets; however, they have a significant limitation in scalability due to a limited training data set \cite{DP_bioNER}. 

In Song et al. and Yan et al., learning representation models based on synonyms and hypernyms were proposed to link biomedical concepts in the literature and social media forums \cite{bib3,bib2}. In addition, unsupervised and self-supervised learning approaches are being explored for concept linking in biomedical literature and social media datasets without task-specific supervision. These methods align the representation spaces of biomedical concepts using scalable metric learning frameworks and have shown promising results on several benchmark datasets of medical concept linking \cite{Zheng2014Entity,liu2021selfalignment}. SapBERT \cite{liu2021selfalignment} refines embeddings for biomedical concepts through self-alignment pre-training, improving the precision of concept identification. Similarly, KRISS BERT \cite{kriss_bert} and BioBERT-snomed \cite{lee2020biobert} have been fine-tuned for biomedical text to enhance entity recognition and linking tasks by incorporating additional context and integrating concept hierarchies, respectively.  

Another work proposed a two-stage concept linking model that used in-context learning to address the challenge of ambiguous mentions \cite{enhancing_entity_rep_prompt_learning_bioent}. The structural information of the knowledge graph has also been utilized through knowledge graph embedding for the concept linking task. In another study, the potential of multilingual bi-encoder models is explored, including various fine-tuned BERT models in the medical domain, highlighting the challenges of linking domain-specific language concepts \cite{Guven2023Multilingual}. 

Most of the aforementioned methods for concept linking face challenges related to generalization and limited training data, leading to decreased accuracy when applied to unseen data during inference. Approaches that used prompt learning and context-aware representation learning mainly focus on free form text, such as clinical notes \cite{enhancing_entity_rep_prompt_learning_bioent,biopro, ehr2vec}. However, recent advances in LLMs, such as GPT \cite{achiam2023gpt} and Llama \cite{touvron2023llama}, have enabled the automation of concept linking for unseen data of different modalities through in-context learning and the integration of external knowledge using RAG \cite{biokg_llm,llm_data_management,entity_match_llm}. In this context, few initiatives have taken advantage of LLMs for concept linking, as demonstrated by \cite{promptlink_MCN} and \cite{incontext_learning_BioNEL}, which specifically target atomic CDEs.

Another work, SPIRES \cite{SPIRES}, proposed structured data extraction from unstructured text and its linking to controlled vocabularies. Although conceptually aligned with our query decomposition methodology, SPIRES focuses on extracting entities and linking atomic concepts from text. However, these solutions prove inadequate when confronted with a vast retrieval space that includes multiple vocabularies and overlapping but semantically distinct concepts. Moreover, these approaches fail to address scenarios where each data element may embody a composite structure, necessitating linking to multiple concepts to establish a cohesive linking.

In contrast, our framework extends these principles to structured data by addressing concept linking for various clinical data representations including atomic CDEs, dependent CDEs and composite CDEs. Our approach is designed to handle linking across multiple controlled vocabularies simultaneously, leveraging RAG system with modular components such as query decomposition, ensemble retrieval, and knowledge reservoir. Given their demonstrated superior performance on concept linking tasks compared to alternative methods, SapBERT \cite{liu2021selfalignment}, KRISS BERT \cite{kriss_bert}, and BioBERT-snomed \cite{lee2020biobert} are widely adopted and well-established baselines for evaluating our approach. 

section{Methods} 
\label{sec3}
The task of linking CDEs to controlled vocabularies involves defining a healthcare data dictionary, denoted \( D \), and a biomedical knowledge base, denoted as \( KB \). The data dictionary \( D \) is structured as table, where each row corresponds to a distinct variable from the underlying data set. The columns in this data dictionary provide metadata for each variable, such as variable label, data type, whether it is continuous/nominal/ordinal, measurement unit, formula, visit and other relevant details \cite{liu2023tabular}. This horizontal structure emphasizes the attributes of each variable, with each row representing a specific variable and the associated metadata describing its properties. The primary objective of this task is to establish links between the elements of \( D \) and the corresponding concepts in \( KB \), thus facilitating the harmonization of the data between various data sources. In this context, clinical data dictionaries refer to metadata describing clinical records, including field names, attributes, and values, which must be linked to controlled vocabularies for standardization.
The \textbf{Data Dictionary \( D \)} can be formally defined as a set of tuples:
\[
D = \{ (M_i, L_i, E_i) \mid i = 1, 2, \ldots, n \}
\]
Each tuple \( (M_i, L_i, E_i) \) contains the following elements:
\begin{itemize}
    \item \( M_i \): The name of a variable.
    \item \( L_i \): The label (description) associated with \( M_i \).
    \item \( E_i \): Additional metadata related to \( M_i \) (e.g., type, measurement units, formula, categories).
\end{itemize}
The \textbf{Knowledge Base \( KB \)} is structured as:
\[
KB = (C, S, T, H)
\]
where:\begin{itemize}
    \item \( C \): Controlled vocabulary terms or concepts e.g. Acute Myocardial Infraction
    \item \( S \): A set of synonyms, each uniquely assigned to a concept in \( C \), e.g., AMI, Myocardia Infarct Actue etc.
    \item \( D \): Semantic types, e.g. medical condition, procedure, etc.
    \item \( H \): Hierarchical relationships between concepts, represented as pairs \( (c_1, c_2) \), where \( c_2 \) is a parent term of \( c_1 \), e.g., Acute myocardial infraction is an acute ischemic heart disease.
\end{itemize}
The goal of concept linking is to identify the terms in \( KB \) that correspond to the elements in \( D \). Specifically, for each label \( L_i \) and \( E_i \) in \( D \), there should exist corresponding description(s) \( C_l \) and \( C_e \) in \( KB \) such that the term(s) in \( C_l \) and \( C_e \) accurately represent \( L_i \) and \( E_i \), respectively.

\begin{table}[ht]
\centering

\begin{tabular}{|p{2cm}|p{2.9cm}|p{2.2cm}|p{2.5cm}|p{3cm}|p{2cm}|}
\hline
\textbf{Dataset} & \textbf{Source} & \textbf{Total Mentions} & \textbf{Unique Concepts} & \textbf{Concepts Type} & \textbf{Composite CDEs} \\ 
\hline
BC5CDR-Disease  & PubMed Articles               & 73,126 & 1,247 & Diseases                            & No  \\ 
\hline
NCBI-DC         & Biomedical Literature         & 73,024 & 359   & Diseases          & No  \\ 
\hline
MIID            & MIMIC-III linked to iBKH-Diseases & 1,493 & 1,493 & Diseases                              & No  \\ 
\hline
HF Studies      & TIME-CHF,CHECK-HF            & 476    & 476   & Demographics, Diseases, Procedures, Visits and Measurements   & Yes \\ 
\hline
\end{tabular}
\caption{Summary of datasets used in the study.}
\label{tab:datasets}
\end{table}

\subsection{datasets} \label{data set}
For this study, we used four diverse datasets with varying characteristics to ensure a comprehensive evaluation and robust results. These datasets include the BC5CDR-Disease corpus \cite{Li2016BioCreative}, the National Center for Biotechnology Information Disease Corpus (NCBI-DC) \cite{NCBI}, the MIMIC-III-iBKH-Disease dataset (MIID) \cite{promptlink_MCN}, and data from Heart Failure (HF) studies, specifically the Trial of Intensified versus Standard Medical Therapy in Elderly Patients with Congestive Heart Failure (TIME-CHF) \cite{maeder2018time_chf} and the Chronisch Hartfalen ESC-richtlijn Cardiologischepraktijk Kwaliteitsproject HartFalen (CHECK-HF) \cite{CardioMEMS}. Ethical approval for CHECK-HF was provided by the Ethical Committee of the Maastricht University Medical Center, the Netherlands, and for TIME-CHF by the Ethical Committee of Both Basel (EKBB), Switzerland. 

We extracted all relevant information from these datasets into a data dictionary using the methodology described in BioSyn \cite{sung2020biomedical}. In particular, only the HF dataset contained composite CDEs, allowing us to assess the effectiveness of our method in handling more complex clinical data structures. Table \ref{tab:datasets} summarizes key characteristics of these datasets, including source, size, annotation type, total entities, unique concepts, and presence or absence of composite CDEs.
Where available, datasets mapped to UMLS Concept Identifiers were further linked to OMOP IDs following the procedure described by \cite{OHDSIananke}. The diverse nature of these datasets, spanning biomedical literature to clinical records, enabled a comprehensive evaluation of our proposed methods across different types of medical concepts.

\subsection{Proposed Framework}\label{subsec3}
\begin{figure*}[htbp]
    \centering
    \includegraphics[width=1\textwidth] {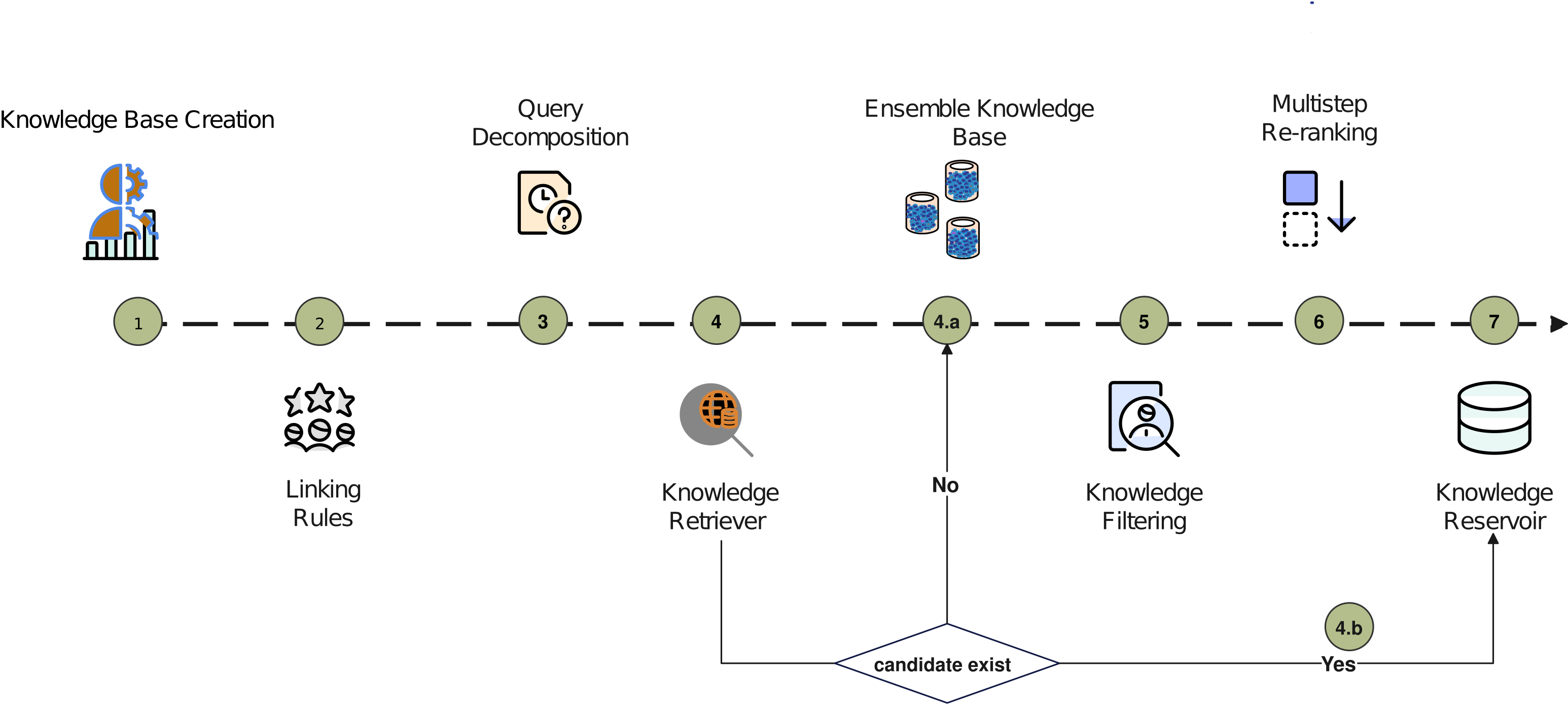}
    \caption{Modular Framework Stepwise Workflow for Concept Linking of Common Data Elements Using RAG}\label{figx}
\end{figure*}

As shown in Fig.\ref{figx}, this overall framework orchestrates a stepwise process incorporating LLMs and a multi-tier retrieval strategy to ensure precise and clinically relevant concept normalization. As described in Section \ref{data set}, we standardize the input format to a data dictionary comprising columns such as description, categories, measurement units, formulas, visits, etc. All inputs are structured in this format, the description of variable is mandatory, while other components are optional. During query decomposition, examples are selected based on contextual relevance to each input. These examples guide the model in performing the decomposition task.

In the next step, query decomposition breaks down composite queries into simpler and manageable components. Then each component is individually processed. In the knowledge retrieval component, the algorithm employs a stack of retrievers to find the most accurate match within a knowledge base. Then a knowledge filtering method is applied to retrieve relevant candidates based on a threshold \( t_s \) for contextually appropriate selection. If an exact match is found, the term is added directly to the result list \( N \). If no exact match is found, the two-step ranking module is triggered to find the most contextually relevant candidate. Finally, correctly predicted concepts are stored in the knowledge reservoir through a human-in-the-loop approach, where experts validate the final mappings to ensure accurate clinical interpretation before retaining them for future use.

\begin{figure*}[htbp]
    \centering
    \includegraphics[width=1\textwidth]{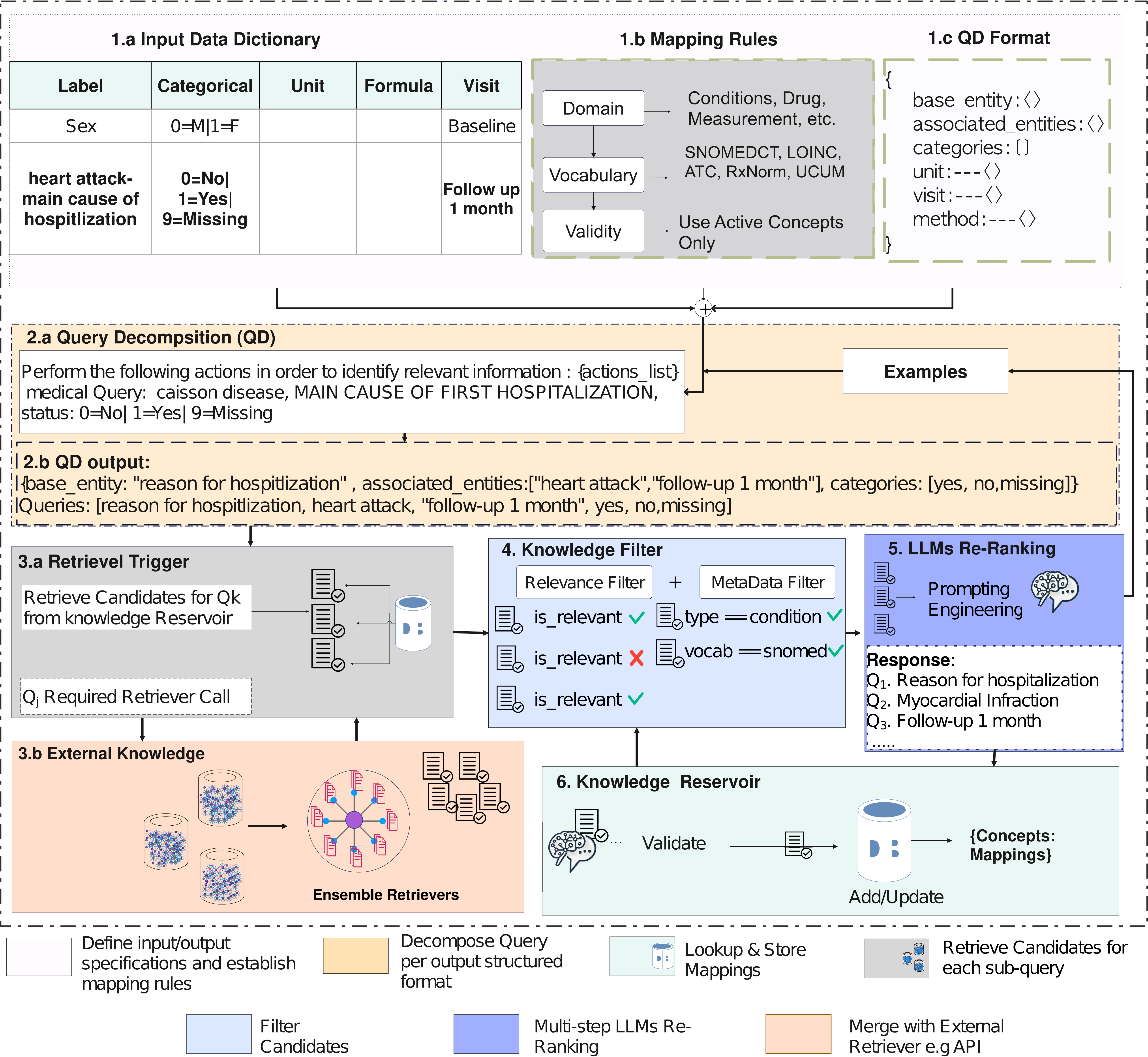}
    \caption{Detailed Overview of Proposed Modular RAG Framework for normalization and linking CDEs in controlled vocabularies (e.g. SNOMED, LOINC, ATC, RxNorm) }
    \label{fig2}
\end{figure*}

To further illustrate the detailed workflow of the framework, Fig. \ref{fig2} provides an example-driven flow chart that showed the stepwise integration of query decomposition, ensemble retrieval, and modular components with input sample data. This detailed representation highlights how the framework standardizes the input format, processes components individually, and links them to unique standard concepts using JSON as the output schema. The structured output (e.g., JSON) with expected schema components, where the base entity, associated concepts, categorical values, measurement units (if available) and so on are individually linked to unique standard concepts. We chose the specific JSON completion format because it provides a standardized hierarchical structure that aligns well with the complexity of clinical common data elements (CDEs). This format ensures consistency in the decomposition of composite queries, supports optional attributes, and simplifies linking to standard terminologies. JSON’s compatibility with downstream systems (e.g., OMOP, FHIR), its interpretability for quality assurance, and its scalability for future extensions make it an optimal choice. 

The process of normalizing medical concepts within the CDE-Mapper framework is detailed in Algorithm in Appendix ~\ref{app:algorithm_section}. It outlines the step-by-step procedure for decomposing queries, applying retrieval methods, and using relevance ranking to ensure precise normalization of healthcare data. Each component is described in detail in the following subsections.

\subsubsection{Knowledge Base Creation}
In this study, we enhance each concept within selected controlled vocabularies by systematically expanding its set of similar terms. First, we remove duplicate or redundant synonyms within the same vocabulary, similar to eliminating stop words in text processing. Next, we incorporate additional synonyms drawn from mapped vocabularies, guided by equivalence relationships (e.g., “Maps to” or “Trade name”). For example, if an ATC concept is related to an equivalent RxNorm concept, RxNorm concept synonyms are merged into the ATC concept terminology set.

Furthermore, in vocabularies with flat hierarchies (e.g., UCUM), the concept codes are aligned with textual labels, so that both forms are recognized as equivalent terms (e.g., mm [Hg] and millimeter mercury column). Table. \ref{table:synonyms} illustrates the representation of the concept in the selected vocabularies before and after the addition of synonyms using this approach. 

This expansion of synonyms offers greater terminological coverage and stronger contextual grounding for downstream tasks of clinical concept linking. By capturing the various ways each concept may appear in clinical or research data, the enriched vocabularies facilitate more accurate mapping and improve overall retrieval performance within embedding-based systems.

\subsubsection{Linking Rules}
The linking rules provide a structured and dynamically configurable approach for linking clinical terms to appropriate source vocabularies based on data types. For example, consider a patient's record indicating a diagnosis of "heart attack/MI" alongside related medications and lab results. The linking rules would be as follows:

\begin{itemize}
    \item Link the condition "heart attack/MI" to a standard vocabulary term from SNOMED CT.
    \item Link prescribed medications, such as "Carvedilol 3mg twice daily," to the corresponding RxNorm identifier. In cases where drug information is less specific, such as "Carvedilol" without dosage details, and a broader classification is sufficient, the ATC classification may be used.
    \item Laboratory measurements, such as "Troponin.T levels," are linked to the appropriate LOINC or SNOMED code.
\end{itemize}

Additionally, contextual factors such as prevalence and incidence are also considered. For example, if a patient's record mentions a "history of myocardial infraction," the system interprets this as part of the patient's baseline information. Consequently, the rule implies mapping this historical context to a specific concept denoting a past condition, rather than an active diagnosis, reflecting its significance in the patient's medical history. In our example, the historical context of the disease informs the concept linking process to differentiate between current and past conditions, preventing misclassification. Employing these flexible rules not only enhances the efficiency of clinical data extraction but also adapts the process to meet specific research requirements. For example, researchers who examine the progression of disease over time can rely on accurately linked concepts to identify patient cohorts, treatment regimens, and outcomes, thereby facilitating more robust and insightful analyzes.

\begin{table}[h!]
\caption{No. of concepts with synonyms per vocabulary}
\resizebox{\textwidth}{!}{%
\centering

\begin{tabular}{|c|c|c|c|}

\hline
\textbf{Vocabulary} & \textbf{Total Concepts} & \textbf{Initial Concepts with Synonyms} & \textbf{Updated Concepts with Synonyms}  \\ \hline
SNOMED & 385573 & 187040 & 381981 \\ \hline
ATC & 6379 & 446 & 6153 \\ \hline
ICD9 & 10648 & 668 & 10406 \\ \hline
ICD10 & 289385 & 19322 & 288409  \\ \hline
LOINC & 216441 & 2861 & 168068 \\ \hline
MeSH & 334573 & 19841 & 22577  \\ \hline
MedDRA & 69041 & 0 & 63556  \\ \hline
RxNorm & 206239 & 445 & 185695  \\ \hline
RxNorm Extension & 1923788 & 0 & 1846835  \\ \hline
UCUM & 1012 & 1 & 1012 \\ \hline
OMOP Extension and Genomics & 382566 & 3349 & 380797  \\ \hline

\end{tabular}
}

\label{table:synonyms}
\end{table}

\subsubsection{Query Decomposition}
The query decomposition module serves two key functions: (1) \textit{enhancing the quality of the original query} and (2) \textit{decomposing the query into a structured completion format}. This process outputs a structured format (e.g., JSON), where each component is treated as a subquery. Formally, it can be expressed as:
\[
G_{\theta}(x) \rightarrow (s, Q)
\]
where $x$ is the original query, $s$ is the refined query, and $Q = \{q_1, q_2, \dots, q_n\}$ represents the set of decomposed sub-queries, with each sub-query $q_i$ corresponding to a specific component of the structured format, such as label, categorical values, unit, method, or formula. The decomposition divides each relevant component into its own subquery. An effective implementation of $G_{\theta}$ leverages an in-context learning prompt-based strategy using task descriptions, original queries, and relevant examples to guide an LLM. 

To address domain-specific data extraction, we collaborated with domain experts to curate high-quality examples, thereby enhancing the robustness and adaptability of the decomposition process. For simple queries with a single variable, the base entity is treated as a standalone sub-query without further decomposition. Consider an original query: 
\textit{``heart attack - main cause of hospitalization, measured at baseline, categories include 0=No, 1=Yes, 9=Missing."}. The decomposition process involves breaking this query into a structured format, such as:

\[
\text{Decomposition Format: } \left\{
\begin{array}{l}
\text{base\_entity: ``heart attack",} \\
\text{associated\_entities: [``Hospitalization Reason"],} \\
\text{categories: [``Yes", ``No", ``Missing"],} \\
\text{visit: ``baseline"}
\end{array}
\right\}
\]

In this example, the \textit{base\_entity} represents the primary medical term (``heart attack"). The \textit{associated\_entities} indicate related concepts such as ``Hospitalization Reason". The \textit{categories} component specifies potential values such as ``Yes, No, Missing".  The \textit{visit} component indicates the timing context, such as the ``baseline."

These components are treated as separate subqueries ($q_1, q_2, q_3$), and individually processed to ensure a precise linking of the concept. It is important to note that in a clinical context, different values for each attribute may be present depending on the scenario.

\subsubsection{Knowledge Retriever}
Our retrieval system utilizes an ensemble candidate retrieval strategy to enhance precision and contextual relevance. The construction of semantic search spaces involves a dual embedding strategy that serves different retrieval needs, depending on the nature of the concepts.

To generate embeddings, we use two complementary models: SPLADE \cite{spladev2} and SapBERT \cite{liu2021selfalignment}. SPLADE is used for concept names where an exact match is expected, referred to as ``canonical concepts." These canonical concepts are standardized and well-defined terms in controlled vocabularies, which are suitable for keyword-based retrieval. SPLADE efficiently represents these terms in a sparse manner, allowing for highly efficient matching by highlighting relevant keywords and reducing ambiguity.

representations using SapBERT, a biomedical-specialized pretrained language model (PLM). The dense embeddings generated by SapBERT are designed to capture subtle variations and relationships between related medical concepts, providing a richer understanding of complex search queries.

Let $\mathcal{C}$ denote the set of queries, and $\mathcal{K}$ denote the set of concepts from the knowledge base. For each query $Q \in \mathcal{C} \cup \mathcal{K}$, the embedding $Q_c$ is generated as follows:
\[
    Q_c = \text{PLM}(c, \text{syn}(c), \text{info}(c)) 
\]
where $\text{syn}(c)$ represents synonyms of $c$, and $\text{info}(c)$ includes hierarchical information, semantic type, etc. Through this hybrid approach to embedding, we aim to establish a search space that can handle both straightforward and complex retrieval. Candidate generation employs the merging retrieval approach, combining the strength of various retriever models. The knowledge retriever module retrieves relevant candidates from both the SapBERT and SPLADE models, which is passed to the knowledge filter module. 

\subsubsection{Knowledge Filter}
LLMs despite their impressive capabilities, are susceptible to generating inaccurate or misleading responses when presented with noisy or irrelevant information retrieved during knowledge recovery. Here, noisy information refers to candidates that are irrelevant, semantically misaligned, or only partially relevant to the query. This noise can arise due to issues such as semantic drift (e.g., retrieving "Fear of heart attack" for a query about "heart attack"), ambiguity in query terms, or false positives caused by over generation during retrieval. To address this critical challenge, we introduce the Knowledge Filter module, designed to enhance accuracy and precision. Although various filtering strategies exist—such as using LLMs for query-candidate classification or leveraging pre-trained language models (PLMs) with similarity measures—our approach focuses on a dual strategy: metadata-based filtering rules and a similarity threshold-based method to improve performance. The Knowledge Filter module discards candidates that significantly deviate from the query. Formally, this process is represented as:
\[
\text{Sim}(E(s), E(k)) < \tau
\]
where $E(s)$ is the embedding of the source candidate, $E(k)$ is the embedding of the query or context, $\text{Sim}()$ calculates the similarity (e.g., cosine similarity) and $\tau$ is the similarity threshold. By filtering out irrelevant or misleading information, the Knowledge Filter reduces the impact of noisy data on LLM output, contributing to more accurate and reliable knowledge-driven responses. The threshold $\tau$ can be adjusted to suit different datasets based on the domain-specific concept linking needs of experts.

\begin{figure*}[htbp]
    \centering
    \includegraphics[width=0.85\textwidth]{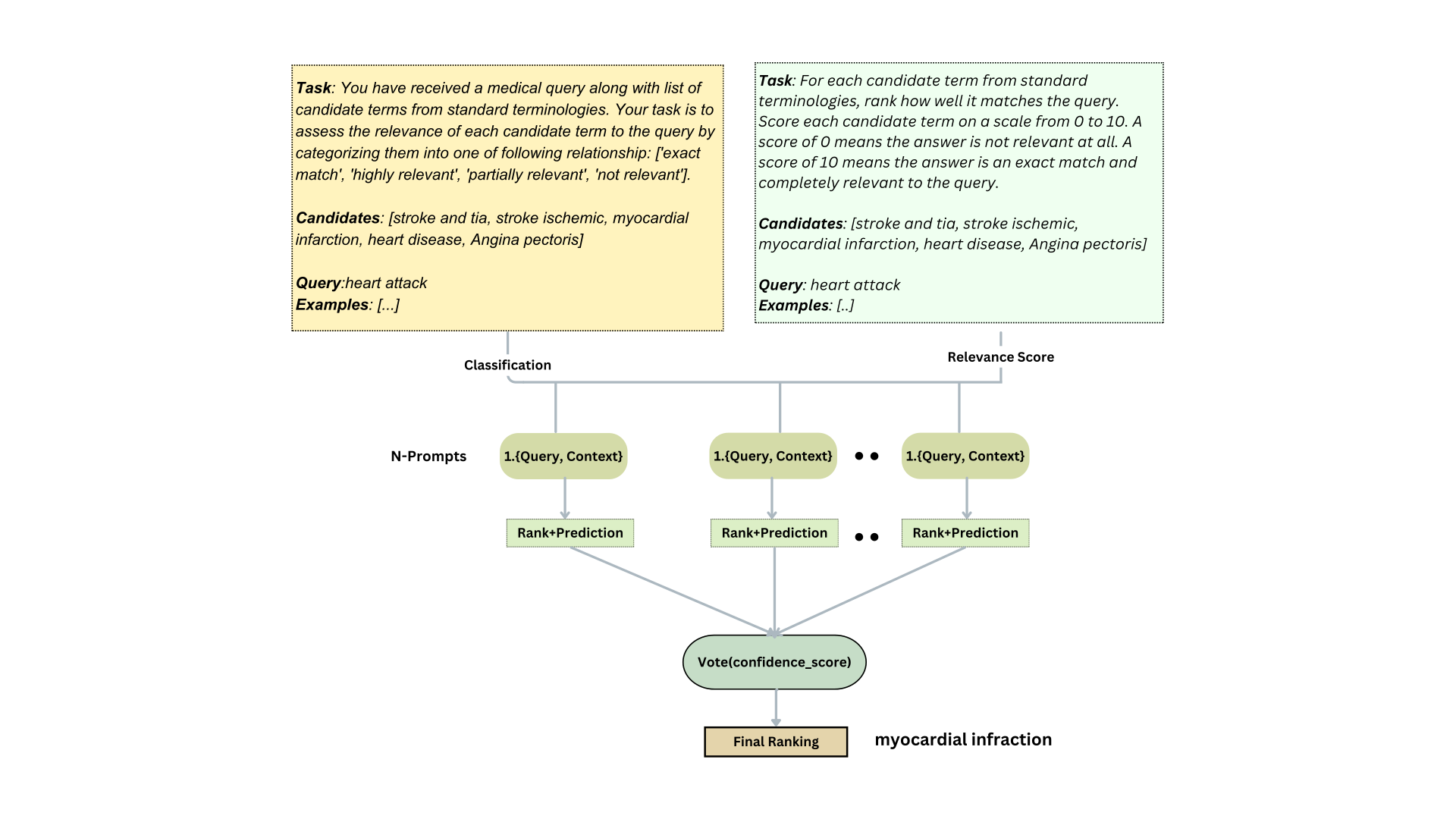}
    \caption{Two-step reranking to select a candidate using a combination of classification and relevance score.}\label{fig3}
\end{figure*}

\subsubsection{Knowledge Reservoir}
The Knowledge Reservoir module aims to reduce the LLM inference time during the search and response generation by storing knowledge retrieved in a structured format. It stores label-concept pairs, where the label is the original query, and concepts consist of relevant candidates along with their concept codes and OMOP IDs. OMOP IDs are utilized because it is possible for two concept codes from different controlled vocabularies to be identical e.g. the ICD9Proc code for ``Other operations in the middle and inner ear" and the OMOP Genomic code for ``measurement of the AARS1 gene variant (alanyl-tRNA synthetase 1)" is 20 and OMOP IDs are 2000982 and 35956407 respectively. Using OMOP IDs ensures better disambiguation and standardization across vocabularies, providing a consistent and reliable reference for concepts. For the knowledge reservoir, two distinct formats can be used for storage, each serving a specific purpose.
\begin{itemize}
    \item \textbf{Dictionary format}: This format supports efficient direct lookup based on exact matches with the original query label. It is particularly useful for frequent, repetitive queries that require rapid responses, as it allows for direct retrieval without the need for complex searching or reasoning.
    
    \item \textbf{Triple store format}: This format uses an ontology to store linked concepts as triples (subject, predicate, object), supporting composite CDEs as queries. The triple store structure is advantageous for storing relationships and contextual information, making it possible to represent hierarchical and semantic relationships between entities. This format is useful for handling more intricate queries that require reasoning over the relationships between various data elements.
\end{itemize}

The reservoir is dynamically updated as new knowledge is retrieved and its schema can be adjusted as needed. Mathematically, it is represented as a set \( K = \{k_1, k_2, \dots, k_{|K|}\} \), where each \( k_i \) is a pair of labels:
\[
k_i = (l_i, \{(c_{i1}, o_{i1}), (c_{i2}, o_{i2}), \dots, (c_{in}, o_{in})\})
\]
Here, \( l_i \) is the label (query), \( c_{ij} \) the $j$th candidate concept, and \( o_{ij} \) the corresponding OMOP code. Initially, an LLM functions as an automated judge, performing zero-shot prompt classification to categorize each concept into  \textit{correct}, \textit{partially correct}, and \textit{incorrect} \cite{llmaszeroshot}. Concepts classified as \textit{correct} or \textit{partially correct} are then subjected to further validation by domain experts. The experts include clinicians who understand the specific datasets, as well as mapping specialists. This expert review ensures that the final mappings are accurate and clinically interpretable before preserving in the Knowledge Reservoir for future use.

\subsubsection{Two-Step Reranking}
As the next step in our framework, we designed a two-step reranking module, as illustrated in Fig.\ref{fig3}. The prompt design includes the task instruction presented in italics for clarity. It is followed by the context, which comprises the selected candidates followed by the query string. This prompt design aligns with the methodological approach described in \cite{liu2024lost}. 

\textbf{Step 1}: Filtered candidates are supplied to a large language model (LLM), which leverages its reading comprehension and reasoning capabilities to evaluate and re-rank candidates on a scale from 1 (lowest) to 10 (highest) based on their relevance to the query.

\textbf{Step 2}: The LLM subsequently classifies the top candidates into predefined relevance categories—namely \textit{not relevant}, \textit{partially relevant}, \textit{highly relevant}, and \textit{exact match} —assigning scores accordingly: 0–4 for not relevant, between 4 and 7 for partially relevant, 8–9 for highly relevant and 10 for exact match. These scores facilitate further processing and selection based on the combined outcomes of the previous steps.

To improve prompt response, we adopt the self-consistency prompting strategy \cite{Lewis2020Retrieval-Augmented}, which involves repeatedly prompting the same question to the LLM(s) multiple times. Specifically, we prompt the query with the top candidates altogether as \( Q_i, [C_i] \) for \( n = 3 \) times. We obtain the confidence score from each response \( B_{i,j} \in [0, 1] \), where the total confidence score is the average of the confidence scores of the multiple prompts. The confidence score for each candidate is determined through a binary transformation of the LLM's relevance scores. Let \( S_{i,j} \) represent the score given to the candidate \( i \) in the prompt \( j \). We define a binary score \( B_{i,j} \) as follows:

\[
B_{i,j} = 
\begin{cases} 
1 & \text{if } S_{i,j} \geq t \\
0 & \text{if } S_{i,j} < t 
\end{cases}
\]

This binary scoring method ensures that only candidates consistently rated as highly relevant across multiple prompts are considered. The confidence score for each candidate \( i \), denoted as \( \bar{B}_i \), is the average of these binary scores on all \( n \) prompts:

\[
\bar{B}_i = \frac{1}{n} \sum_{j=1}^{n} B_{i,j}
\]

A candidate is deemed relevant if its average confidence score \( \bar{B}_i \) exceeds a predefined threshold \( \tau \). The threshold \( \tau \) is set to \( 0.85 \times n \), where \( n \) is the number of prompts:

\[
\tau = 0.85 \times n
\]

Thus, a candidate \( i \) is considered relevant if \( \bar{B}_i > \tau \). Based on our preliminary work (results not shown) to determine the appropriate threshold, we observed that a higher threshold is required in cases where multiple closely aligned candidates are present. This strategy effectively prioritizes the most relevant match among closely aligned candidates while filtering out candidates with lower scores, which are deemed irrelevant. This rigorous filtering criterion ensures that only those candidates that consistently achieve high relevance scores in multiple evaluations are selected. We consider a candidate highly relevant if it consistently appears among the top-ranked candidates on multiple prompts. For example, if the LLM identified the same top 3 candidates as highly relevant in each of the prompts, their confidence scores reinforce their relevance. In case that all candidates are irrelevant, ``NA'' is returned.

\section{Experiment Design} \label{4}

This study compares our approach to linking the biomedical concept with several state-of-the-art models, including SapBERT \cite{liu2021selfalignment}, KRISS BERT \cite{kriss_bert}, BioBERT-snomed \cite{lee2020biobert}, and the LLM-based approach PromptLink \cite{promptlink_MCN}. These baseline models were selected for their demonstrated effectiveness in biomedical tasks and their emphasis on linking individual medical terms to controlled vocabularies. The key differences between our approach and these are that they do not perform retrieval and relevance reranking for composite CDEs.

The datasets presented in Table. \ref{tab:datasets} were used as test entities / variables for concept linking to assess the performance of the proposed approach and the baseline methods. This ensures a consistent and unbiased comparison across all models. Importantly, these datasets represent diverse biomedical contexts, encompassing a mix of controlled vocabulary terms, composite terms, thereby providing a comprehensive benchmark for evaluation.

For query decomposition, we used GPT4 and GPT4o-mini (closed source) and Llama3.1 70b (open source) as black-box LLM. We used SapBERT \cite{liu2021selfalignment} and the SPLADE model \cite{spladev2} to generate a semantic search space in the knowledge retrieval module. For similarity, we used the consine similarity alogirthm. For optimal retrieval, the scalar quantization method is employed in the vector database (qdrant v1.10). For the knowledge retrieval module, we selected k=10 as the number of top candidates to retrieve. This decision was based on several considerations. First, k = 10 is an accepted baseline and is in agreement with standard practices in similar tasks, where values in the range of 10–20 are commonly used \cite{liu2021selfalignment,kriss_bert,promptlink_MCN,lee2020biobert}. This range balances retrieval efficiency and accuracy, including the most relevant candidates without overwhelming the reranking module. For LLM reranking module, we utilized the same models as utilized in query decomposition module. To ensure a fair comparison with existing methods, we employed accuracy as a primary metric, as it directly reflects the performance of baseline models on similar tasks, particularly in evaluating the accuracy of the top-ranked predictions. However, since our framework incorporates a reranking module to refine the matched candidate, we also used the Normalized Cumulative Gain Difference (NCGD) metric. NCGD provides a nuanced evaluation of the ranking performance considering the relevance and order of retrieved candidates, making it particularly suitable to assess the effectiveness of our ranking approach in prioritizing clinically relevant matches \cite{ncgd_ir}.

\begin{table}
    \caption{Performance comparison of CDE-Mapper variants and baseline models on concept linking tasks. datasets include NCBI-DC, BC5CDR-D, and HF Studies for linking to controlled vocabularies, as well as MIID for linking MIMIC-IV concepts to IBKH knowledge base. Results are presented as accuracy at top-1 (acc@1). - Denotes unavailable results from baseline.  Statistically significant improvements (†) are noted compared to both baseline models and other CDE-Mapper variants using different LLMs (T-test, $\rho$ < 0.05).}

    \begin{tabular}{|l|r|r|r|r|}
    \hline
    \textbf{Models}  & \textbf{NCBI-DC} & \textbf{BC5CDR-D} & \textbf{MIID} & \textbf{HF Studies} \\ \hline
    SapBERT & 83.2 & 86.2 & 72.9 & 77.2 \\ \hline
    KRISS BERT & 81 & 83 & 50.7 & 58.9\\ \hline
    BioBERT-snomed &80.1 & 84.8 & 64.5 & 39.2\\ \hline
    PromptLink & - & - & 77.5 & -   \\ \hline
    CDE-Mapper(Llama3.1) & \textbf{94.4}† & 88.2 & \textbf{80.2}† &\textbf{86.4}†\\ \hline
    CDE-Mapper (GPT4o-mini) & 93.3 & \textbf{88.6}† & 79.2 & 85.2\\ \hline
    CDE-Mapper (GPT4) & 88.1 & 88.3 & 79.3 & 83.2\\ \hline
    
    \end{tabular}
    \label{tab:model_comparison}
\end{table}
\section{Results}\label{sec5}
Our task focuses on automating the concept linking of CDEs using RAG models. The evaluation of various models across datasets, as shown in Table \ref{tab:model_comparison}, highlights the performance of each approach. Our methodology, leveraging Llama3.1, GPT4o-mini, and GPT4, consistently outperforms baseline models, demonstrating superior capability in handling various types of CDEs.

\subsection{Performance Compared to Baselines}

Baseline models, including SapBERT, KRISS BERT, and BioBERT-snomed, exhibited high performance on datasets rich in disease mentions, such as NCBI-DC and BC5CDR-D. For example, SapBERT achieved accuracy of 83. 2\% in NCBI-DC and 86. 2\% in BC5CDR-D. In comparison, CDE-Mapper using Llama3.1 achieved the highest accuracy of 94.4\% in NCBI-DC, outperforming SapBERT by more than 11\%. In the BC5CDR-D dataset, GPT4o-mini reached an accuracy of 88.6\%, exceeding baseline performance.

In the MIID dataset, Llama3.1 showed a 2.7\% improvement over the baseline models. In the HF Studies dataset, which includes composite CDEs, baseline models struggled. SapBERT achieved an accuracy of 77\%, and KRISS BERT reached 58.9\%. In contrast, CDE-Mapper using Llama3.1 demonstrated a significant improvement, achieving an accuracy of 86.4\%.

\begin{figure}[ht]
    \centering
    \begin{subfigure}[b]{0.5\textwidth}
        \includegraphics[width=\textwidth]{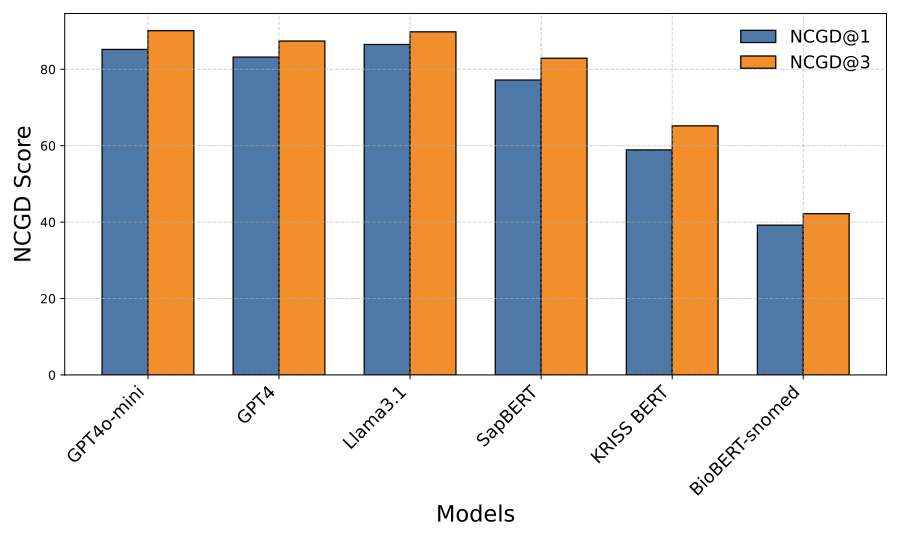}
        \caption{NCGD@K  for HF Studies }
        \label{fig:ncgd_hf_studies}
    \end{subfigure}
    \hfill
    \begin{subfigure}[b]{0.49\textwidth}
        \includegraphics[width=\textwidth]{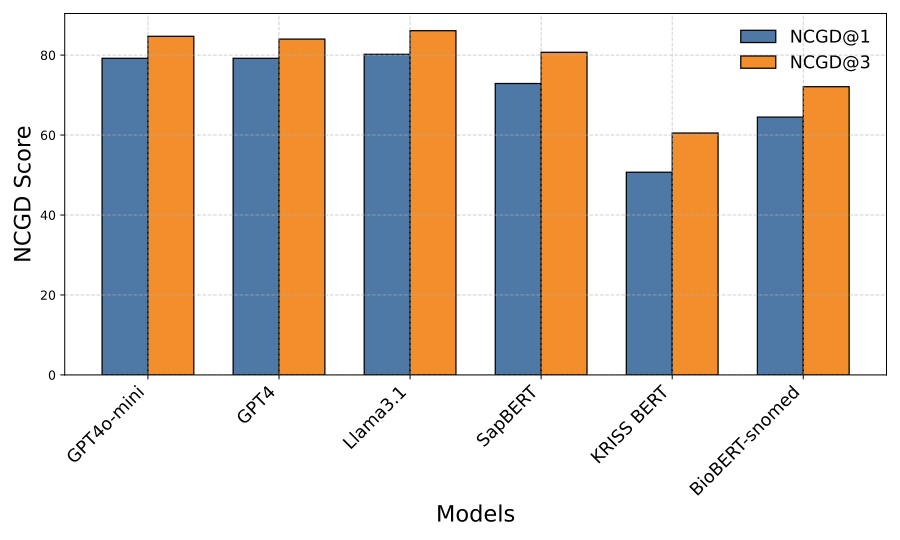}
        \caption{NCGD@K for MIID}
        \label{fig:ncgd_miid}
    \end{subfigure}
    \vskip\baselineskip
    \begin{subfigure}[b]{0.5\textwidth}
        \includegraphics[width=\textwidth]{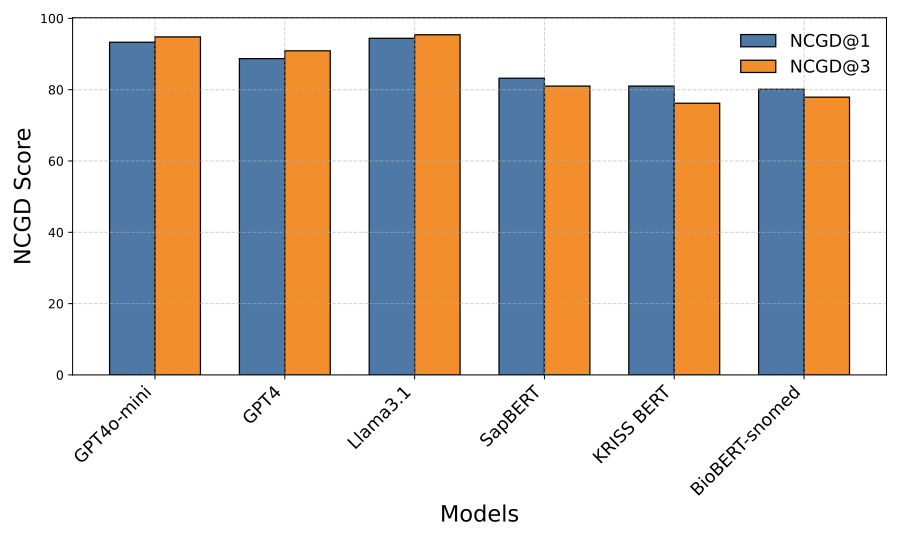}
        \caption{NCGD@K for NCBI-DC}
        \label{fig:ncgd_NCBI}
    \end{subfigure}
    \hfill
    \begin{subfigure}[b]{0.49\textwidth}
        \includegraphics[width=\textwidth]{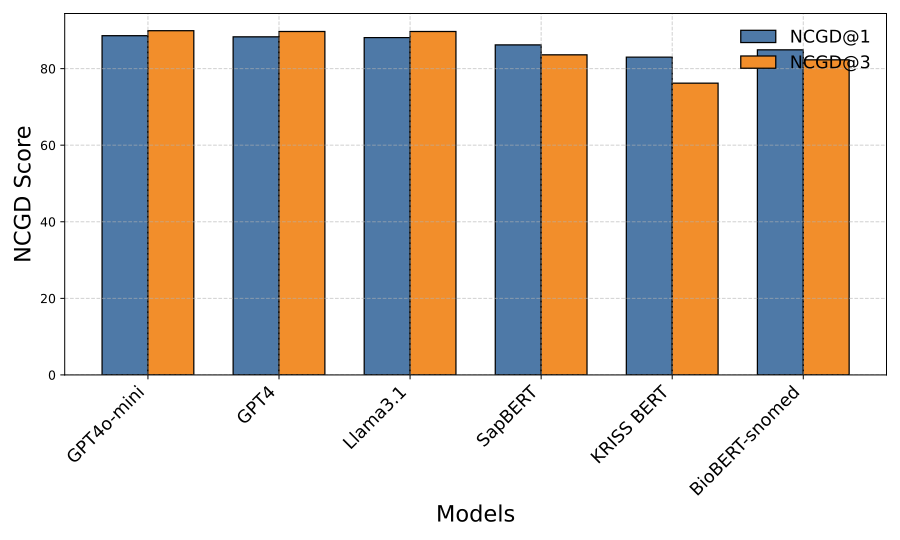}
        \caption{NCGD@K  for BC5CDR-D}
        \label{fig:ncgd_bc5cdrd}
    \end{subfigure}
    \caption{Comparison of NCGD@K  for Different Models Across datasets}
    \label{fig:all_ncgd_scores}
\end{figure}

To further evaluate the effectiveness of the CDE-Mapper, we used the NCGD@K metric, which provides additional insight into the model's ranking quality across various datasets. As depicted in Fig.\ref{fig:all_ncgd_scores}, the CDE-Mapper showed consistent performance at the NCGD@1 and NCGD@3 levels, outperforming the baseline models. Baseline models such as BioBERT-snomed and KRISS BERT showed a noticeable decline in NCG@k values across different datasets, particularly for HF and MIID datasets. This indicates their limitations in capturing domain-specific contextual relationships, which are crucial to handling complex clinical terminologies. For example, in the HF Studies , we observed that BioBERT-snomed struggles to rank the correct concepts high enough, leading to a steeper drop in the NCG@k values.

\subsection{Ablation Studies and Component Analysis}
To further understand the contributions of individual components within our CDE-Mapper framework, we performed ablation studies across all datasets for various types of CDE. The results, summarized below, illustrate the impact of each component on overall performance.

\begin{table}[]
\caption{Accuracy (Acc@1) of CDE-Mapper and baseline models (SapBERT, KRISS BERT, BioBERT-snomed) across different CDE types. The CDE-Mapper results are further divided into three variants: Llama3.1, GPT4o-mini, and GPT4. CDEs types include atomic CDEs , composite CDEs which further categories into dependent CDEs. This comparison highlights the performance of each model in linking CDEs to controlled vocabularies, with CDE-Mapper (Llama3.1) consistently achieving the highest accuracy.}
\begin{tabular}{|c|ccc|c|c|c|}
\hline
\multirow{2}{*}{CDEs Type} & \multicolumn{3}{l|}{CDE-Mapper}                                       & \multirow{2}{*}{SapBERT} & \multirow{2}{*}{KRISS BERT} & \multirow{2}{*}{BioBERT-snomed} \\ \cline{2-4}
                          & \multicolumn{1}{l|}{Llama3.1} & \multicolumn{1}{l|}{GPT4o-mini} & GPT4 &                          &                             &                                 \\ \hline
Atomic CDEs                      & \multicolumn{1}{l|}{\textbf{89.7}}     & \multicolumn{1}{l|}{87.6}       & 88   & 83.8                     & 65.9                        & 50.3                            \\ \hline

Composite CDEs                      & \multicolumn{1}{l|}{\textbf{80.5}}     & \multicolumn{1}{l|}{\textbf{80.5}}       & 71.5 & 65.4                     & 45.1                        & 18                              \\ \hline
Dependent CDEs                      & \multicolumn{1}{l|}{80}       & \multicolumn{1}{l|}{80}         & \textbf{86.7} & 66.7                     & 53.3                        & 30                              \\ \hline
\end{tabular}

\label{tab:cde_comparison_split}
\end{table}

\subsubsection{Performance by CDE Type}
To evaluate the performance of CDE-Mapper in various types of CDE, we assessed accuracy at (acc@1) compared to baseline models such as SapBERT, KRISS BERT and BioBERT-snomed, as shown in Table. \ref{tab:cde_comparison_split}. 

The analysis highlights the ability of CDE-Mapper to handle atomic CDEs (atomic CDEs), dependent CDEs (dependent CDEs), and composite CDEs (composite CDEs) effectively, even when challenged with varying levels of complexity and contextual nuances.

For atomic CDEs, which represent singular and straightforward data elements (e.g., patient sex or age), CDE-Mapper achieved the highest precision of 89. 7\%, outperforming SapBERT (87. 6\%) and KRISS BERT (88\%). This reflects its strong ability to standardize atomic terms consistently.

Composite CDEs involve complex and interconnected data elements, such as medical conditions with multiple attributes such as severity, onset, and related procedures. For these CDEs, CDE-Mapper achieved an accuracy of 80.5\%, significantly outperforming KRISS BERT (71.5\%) and BioBERT-snomed (65.4\%). In the case of dependent CDEs, where normalization depends heavily on the context (e.g. biomarkers such as NT-proBNP, which differ in meaning depending on whether measured in plasma or serum), the CDE-Mapper (Llama3.1, GPT 4o-mini) demonstrated notable performance with an accuracy of 80 5\%, significantly exceeding SapBERT 66. 7\%.

\begin{table}[ht]
   
    \centering
     \caption{Query Decomposition on HF Studies data set}
    \begin{tabular}{|l|r|r|r|}
        \hline
        \textbf{Model} & \textbf{Metric} & \textbf{Attribute Extraction} & \textbf{Value Extraction} \\ \hline
        \multirow{3}{*}{Llama3.1} & Precision & $0.63$ & $0.83$ \\ 
                                  & Recall    & $1.00$ & $0.86$ \\ 
                                  & F1        & $\textbf{0.78}$ & $0.84$ \\ \hline
        \multirow{3}{*}{GPT4} & Precision & $0.63$ & $0.85$ \\ 
                              & Recall    & $1.00$ & $0.84$ \\ 
                              & F1        & $\textbf{0.78}$ & $\textbf{0.85}$ \\ \hline
        \multirow{3}{*}{GPT4o-mini} & Precision & $0.63$ & $0.83$ \\ 
                                    & Recall    & $1.00$ & $0.84$ \\ 
                                    & F1        & $\textbf{0.78}$ & $0.82$ \\ \hline
    \end{tabular}
    
    \label{table:query_decomposition_results}
\end{table}

\subsubsection{Performance of Query Decomposition} \label{query_decompisiton}
We evaluated the query decomposition module using the data set from the HF Studies, where the definitions of the variables and associated attributes served as input. The ground truth was manually curated with the help of two domain experts. The evaluation of the true positive was conducted using two key metrics:

\begin{itemize}
    \item \textbf{Exact word-match}: Measures how accurately the model extracted attributes (e.g., base entity, domain, associated entities, categories, unit, etc.).
    \item \textbf{Fuzzy word-match}: Measures how accurately the model extracted values for each relationship, allowing minor variations. 
    \item \textbf{Base entity}: The mention is a base entity or is grouped with a valid base entity or it is in the correct field in the correct base entity group (Json-Object)
\end{itemize}

The results of query decomposition are summarized in Table~\ref{table:query_decomposition_results}. All models demonstrated identical precision in attribute extraction, each achieving a precision of 0.63 and a perfect recall of 1.00, resulting in an F1 score of 0.78. In value extraction, the models maintained similar precision scores, ranging from 0.83 to 0.85. However, slight variations in recall led to differences in F1 scores. GPT4 achieved the highest F1 score of 0.85 due to its superior precision (0.85) and competitive recall (0.84), while Llama3.1 achieved an F1 score of 0.84 with a precision of 0.83 and a recall of 0.86. GPT4o-mini recorded a slightly lower F1 score of 0.82, maintaining a precision of 0.83 and a recall of 0.84. These findings indicate that, while precision was consistent across all models, minor differences in recall contributed to variations in their overall performance in value extraction.

\begin{table}[ht]
 \caption{Acc@k Retrieval Performance with Different Retriever Configurations Across datasets}
    \centering
    \begin{tabular}{|l|l|r|r|r|r|}
    \hline
    \textbf{datasets} & \textbf{Retriever} & \textbf{Acc@1} & \textbf{Acc@3} & \textbf{Acc@5} & \textbf{Acc@10} \\ \hline
    \multirow{4}{*}{MIID} 
        & Ensemble (label only) & 70.3 & 86.2 & 89.4 & 90.7 \\ 
        & Ensemble+KF (label only) & 71.5 & 88.1 & 90 & 90.8\\ 
        & Ensemble (context-aware) & 73.3 & 86.2 & 89.4 & 91 \\ 
        & Ensemble+KF (context-aware) & 75.3 & 88 & 90 & \textbf{92.8}  \\ \hline
        
    \multirow{4}{*}{NCBI-DC} 
        & Ensemble (label only) & 91.5 & 94.6 & 95.2 & 95.2 \\ 
        & Ensemble+KF (label only) & 88.7 & 91.8 & 92.4 & 92.6 \\
        & Ensemble (context-aware) & 92.7 & 95.8 & 96.3 & 96.6 \\ 
        & Ensemble+KF (context-aware) & 92.7 & 95.8 & 96.3 & \textbf{96.6} \\ \hline
        
    \multirow{4}{*}{BC5CDR-D} 
        & Ensemble (label only) & 90.5 & 94 & 95.2 & 95.9 \\ 
        & Ensemble+KF (label only) & 90.5 & 94 & 95.2 & 95.9 \\ 
        & Ensemble (context-aware) & 90.5 & 94.3 & 95.2 & 95.9 \\ 
        & Ensemble+KF (context-aware) & 90.5 & 93.6 & 95 & \textbf{95.9} \\ \hline
        
    \multirow{4}{*}{HF Studies } 
        & Ensemble (label only) & 76.9 & 91.4 & 94.5 & 95.8  \\ 
        & Ensemble+KF (label only) & 77.4 & 91.6 & 94.5 & 95.8 \\ 
        & Ensemble (context-aware) & 78.1 & 91.6 & 94.5 & 95.8 \\ 
        & Ensemble+KF (context-aware) & 77.8 & 91.4 & 94.7 & \textbf{96} \\ \hline
    \end{tabular}%
   
    \label{tab:retriever_topk_analysis}
\end{table}

\subsubsection{Impact of Knowledge Filtering (KF)}
The introduction of knowledge filtering, combined with a merger retriever that integrates dense and sparse retrieval mechanisms, resulted in significant improvements in retrieval accuracy. Table. \ref{tab:retriever_topk_analysis} presents the retrieval performance with accuracy at k (acc@k) of various retrieval configurations in selected datasets. The results indicate that the incorporation of KF consistently improved the retrieval performance in most datasets. Specifically, transitioning from \textbf{Ensemble (label only)} to \textbf{Ensemble+KF (label only)} yields modest improvements in certain datasets such as MIID, with acc@1 increasing from 70. 3\% to 71. 5\%. Additionally, context-aware configurations showed further benefit, demonstrating higher retrieval accuracies compared to the non-KF counterparts. The \textbf{Ensemble+KF (context-aware)} configuration consistently outperforms other MIID configurations, achieving an acc@1 of 75.3\%, highlighting the synergistic benefits of combining ensemble methods with knowledge filtering and context-awareness. For the HF Studies data set, \textbf{Ensemble+KF (context-aware)} demonstrated slight improvements in acc@1, increasing from 76\% to 77.8\%.

\subsubsection{Role of Two-Step Reranking}
To evaluate the effectiveness of the stepwise reranking mechanism within our framework, we performed additional experiments on selected datasets using different models. Specifically, we evaluated the performance implications of both steps in our two-step reranking process. The first step involves reranking candidates using a relevance score ranging from 1 to 10, while the second step classifies these re-ranked candidates into predefined relevance categories based on their scores.

The reranking performance for Llama3.1, GPT4o-mini, and GPT4 models across various datasets at acc@[1,3] is shown in the Table. \ref{tab:reranking_performance}. The results demonstrate that reranking consistently improvec retrieval performance in all datasets and models. For example, in the HF Studies data set, Llama3.1 improvec from 85.4\% at step 1 to 86.2\% at step 2, while GPT4o-mini showed an improvement from 84.5\% to 86.4\%. Similarly, GPT4 improved from 81.4\% at step 1 to 83.2\% at step 2, indicating that the reranking step effectively enhances the retrieved results.

In particular, the NCBI-DC data set showed the highest performance gains, with Llama3.1 reaching 94. 4\% in step 2. In contrast, the MIID data set showed modest improvements across all models, likely due to the inherent complexity and variability of the concepts involved. The BC5CDR-D data set maintained stable performance, with GPT4o-mini reaching the highest accuracy of 88.6\%.

\begin{table}[ht]
 \caption{Reranking at Acc@[1,3] Performance of LLMs}
    \centering
    \begin{tabular}{|c|c|c|c|c|}
    \hline
    \textbf{Models} & \textbf{HF Studies} & \textbf{NCBI-DC} & \textbf{MIID} & \textbf{BC5CDR-D} \\ \hline
    Llama3.1 (Step 1) & 85.4, 89 & 94.2, 96.4 & 79.9, 89.3 & 87.1, 91.9 \\ \hline
    Llama3.1 (Step 2) & \textbf{86.4}, 92.3 & \textbf{94.4}, 96.6 & \textbf{80.2}, 89.8 & 88.1, 92.8\\ \hline
    GPT4o-mini (Step 1) & 84.5, 93.2 & 92.8, 96.4 & 70.8, 75.4 & 87.9, 93.1 \\ \hline
    GPT4o-mini (Step 2) & 85.2, 93.4 & 93.3, 96.9 & 79.2, 88.1 & \textbf{88.6}, 90.3 \\ \hline
    GPT4 (Step 1) & 81.4, 90.2 & 87.5, 92.4 & 77.5, 86.6 & 87.3, 92.6 \\ \hline
    GPT4 (Step 2) & 83.2, 91.7 & 88.1, 92.4 & 79.3, 86.8 & 88.3, 92.6 \\ \hline
    \end{tabular}%
    \label{tab:reranking_performance}
\end{table}

\subsection{Case Study}
To further assess the performance of our proposed CDE-Mapper framework, we conducted a case study using various CDEs is shown in the Table ~\ref{tab:model-predictions}. This analysis focuses on instances where baseline models such as SapBERT, KRISS BERT, and BioBERT-snomed either struggled or succeeded in comparison to CDE-Mapper, particularly in handling complex CDEs, rare conditions, and ambiguous terminologies.

In this case study, we evaluated three scenarios including (1) concepts evaluated by both ground truth labels and two cliniciasn; (2) concepts evaluated by clinicians due to missing ground truth labels; (3) cases that require further context. Overall, the CDE-Mapper model showed consistently higher performance in all three cases compared to baseline. 
In Case II, the unit ``pmol/L" (picomole per liter) is a precise measurement used in laboratory results. SapBERT and KRISS BERT incorrectly mapped this to ``micromole per liter", which represents a different magnitude and could lead to significant clinical misinterpretations. BioBERT-snomed mapped it to ``picogram per milliliter", also an incorrect unit. In contrast, CDE-Mapper accurately linked it to ``picomole per liter", matching the ground truth and demonstrating its ability to handle specific measurement units accurately.

In scenarios without ground truth, such as Case XI involves the composite CDE ``heart rate measured in the recumbent position", where the positional context is clinically significant. SapBERT mapped this to ``resting heart rate", and KRISS BERT to ``pulse rate measured by palpation", both of which failed to capture the importance of the recumbent position. BioBERT-snomed provided ``heart rate unspecified time mean by the pedometer", which is unrelated. CDE-Mapper successfully mapped it to joint concepts``heart rate | position of the occupant body", effectively capturing the positional context necessary for an accurate clinical evaluation.

Another challenging example is Case XII, which involves ``mild to moderate asthma exacerbations". This term requires an accurate standardization of both the condition and its severity. SapBERT mapped it to ``exacerbation of mild persistent asthma", capturing the "mild" aspect but potentially missing the ``moderate" severity. KRISS BERT mapped it to ``moderate persistent asthma", focusing on the ``moderate" aspect. BioBERT-snomed also mapped it to ``exacerbation of mild persistent asthma". CDE-Mapper mapped it to two concepts ``acute asthma|mild to moderate", encompassing both severity levels and aligning with clinical interpretations. This demonstrates CDE-Mapper's capability to handle composite terms involving nuanced severity gradations more effectively than the baseline models.

In Case XIV, the term ``geniculate herpes zoster" refers to a specific viral infection affecting the facial nerve, commonly known as herpes zoster oticus or Ramsay Hunt Syndrome Type II. SapBERT correctly mapped it to ``herpes zoster oticus", aligning both with ground truth and clinical acceptance. KRISS BERT incorrectly assigned it to ``herpes simplex keratitis", which is a different condition affecting the eye. BioBERT-snomed mapped it to a synonym term ``herpes zoster auricularis", a clinically correct interpretation. CDE-Mapper mapped it to ``Ramsay Hunt Syndrome Type II", another synonym, while clinically appropriate, is not the correct term according to the ground truth. This indicates that CDE-Mapper is able to find clinically relevant terms where ground truth may or may not include them, which is crucial for clinical accuracy and interoperability.

However, in Case VII, involving ``nonspecific T wave abnormality, improved on ECG'', none of the models, including CDE-Mapper, produced the correct candidate. Furthermore, in Case XIV, while CDE-Mapper mapped ``geniculate herpes zoster'' to a clinically relevant term, it did not align with the exact standardized terminology expected according to the ground truth and clinicals. These instances highlight areas where CDE-Mapper excels and areas that require further refinement.

\section{Discussion}\label{sec6}
The empirical results demonstrate the efficacy of the CDE-Mapper framework compared to established baseline models. Specifically, CDE-Mapper, which uses Llama3.1 and GPT4o-mini, consistently outperforms these baselines in multiple datasets. This superior performance suggests that the integration of RAG techniques, knowledge filtering, and a two-step reranking process significantly enhances concept linking accuracy. 

For query decomposition, overall all models demonstrated comparable performance in attribute extraction, achieving perfect recall but moderate precision. For value extraction, both Llama3.1 and GPT4 achieved high recall, with GPT4 excelling in precision. Despite these strengths, all models exhibited similar weaknesses, particularly in normalizing context-dependent measurements, units, and abbreviations or acronyms. These observations highlight the potential for improvement in handling nuanced clinical data elements requiring domain-specific contextual understanding.

Incorporation of knowledge filtering significantly improved performance in the MIID dataset, with the Ensemble+KF (context-aware) configuration achieving an acc@1 of 75.3\%. In the BC5CDR-D dataset, no significant improvements were observed, with \textbf{Ensemble+KF (context-aware)} achieving performance similar to other configurations. This suggests that the limited enhancement is due to the scarcity of contextual information available for many concepts in the data set.  Of 73,412 concepts, only 6,530 had hierarchical information (parent terms), and 7,097 had synonyms. The lack of sufficient synonymic and hierarchical data constrained the effectiveness of knowledge filtering and context-aware methods, underscoring the reliance of our approach on the availability of comprehensive and faceted attributes. Despite some datasets showing limited improvement, our findings demonstrate the overall effectiveness of Knowledge Filtering in refining retrieval results, improving both precision and relevance. In addition, the integration of context-sensitive mechanisms, which leverage semantic information, contributed to improved performance.

Similarly, the two-step reranking mechanism effectively refined retrieval results: models such as Llama3.1, GPT4o-mini, and GPT4 showed measurable improvements from step 1 to step 2 in various datasets. The findings suggest that the NCBI-DC dataset benefits the most from reranking, while more complex datasets like MIID showed modest gains, possibly due to varying labels and definitions of entities. The BC5CDR-D dataset, with limited contextual synonyms and hierarchical information, demonstrated minimal improvement with knowledge filtering, implying a dependency on rich semantic data for optimal performance. It is also observed that reranking did not yield significant performance improvements in some cases, as acc@k where k in [5,10] in knowledge filtering was much higher than step 2 of reranking. 

Each component contributes incrementally to the overall accuracy, highlighting the importance of a multifaceted approach in the standardization of composite CDEs. These findings provide the value of combining retrieval methods with robust filtering and reranking processes to achieve superior concept linking results. In the knowledge reservoir, the dual layer validation process, combining automated LLM assessment with expert oversight, prevents the propagation of erroneous or suboptimal concepts, thereby upholding the integrity and quality of the Knowledge Reservoir. Only concepts that pass both automated and expert validations are retained, ensuring that the knowledge base remains accurate and reliable for downstream applications.

The integration of context-sensitive mechanisms, such as those that utilize semantic information from diverse controlled vocabularies, contributed to improved performance and more precise concept linking. In general, the discussion of these results highlights a clear pathway to enhanced retrieval performance through a multifaceted approach, emphasizing the value of combining retrieval methods with robust filtering and re-ranking processes.

\subsection{Limitations}
This study demonstrates the significant potential of leveraging RAG models to map atomic CDEs, dependent CDEs, and composite CDEs to the standardized vocabularies, overcoming the challenges faced by existing methods. However, several limitations highlight areas for future improvement.

One major limitation lies in the variability of model outputs based on the quality of contextual examples provided to the LLMs. Although the default linking rules template is robust, deviations in template structure or contextual examples can lead to inconsistently linked concepts. This highlights the dependency on domain expertise to create high-quality prompts and examples. Future research should explore automated techniques for generating effective prompts, such as reinforcement learning from human feedback (RLHF) or low-resource fine-tuning methods.

Performance gains through reranking were not always highly significant, suggesting the need to refine the reranking mechanism or explore alternative methods to ensure the retrieval of highly relevant results. Furthermore, some edge cases, particularly those involving nuanced contextual linking (e.g. ``geniculate herpes zoster", ``nonspecific T wave abnormality, improved on the ECG"), highlight the limitations of RAG models. Such cases emphasize the need for continuous model refinement and the incorporation of human oversight, especially in high-stakes clinical applications.

Another limitation is the framework's reliance on comprehensive and high-quality controlled vocabularies. The success of knowledge-filtering and-retrieval components depends on the availability of rich synonymic and semantic relationships in the knowledge base. Sparse or incomplete vocabularies may limit the framework's efficacy, as observed in datasets like BC5CDR-D, where insufficient synonyms constrained performance improvements. Further research could investigate the dynamic integration of multiple knowledge sources or the addition of existing vocabularies with crowd-sourced data.

Finally, the computational cost and inference speed of LLMs remain significant challenges. Although open source models such as Llama3.1 offer a cost-effective alternative to proprietary models such as GPT-4, they still require substantial computational resources for local deployment. Techniques such as model quantization, distillation, and adaptive compression could reduce the computational footprint while maintaining accuracy. Furthermore, developing domain-specific LLMs with efficient architectures could address the high-resource requirements for inference.

\section{Conclusions} \label{sec7}

This study introduces CDE-Mapper, a novel framework that uses the RAG technique to automate the linking of clinical data elements to controlled vocabularies. By integrating modular components for query decomposition, ensemble retrieval, knowledge filtering, and two-step reranking, our approach demonstrates significant advancements over baseline models, particularly in handling complex and composite CDEs.

Our results across diverse datasets showed the framework's scalability, achieving state-of-the-art accuracy while preserving cost-effectiveness through open-source models like Llama3.1. These improvements are critical to improving data interoperability, enabling seamless integration of heterogeneous datasets, and supporting advanced clinical research and decision making.

Future research should focus on addressing identified limitations, such as optimizing computational efficiency, expanding vocabulary coverage, and refining contextual understanding. By incorporating adaptive learning techniques and domain-specific enhancements, the framework can further improve its robustness and applicability to diverse healthcare contexts. Ultimately, this work highlights the transformative potential of combining RAG models with controlled vocabularies, paving the way for more accurate, scalable, and efficient data standardization solutions in the era of digital health.

\clearpage 
\begin{landscape}
\begin{table*}[htbp]
    \centering
    \begin{threeparttable} 
        \footnotesize 
        \begin{tabular}{|c|P{3cm}|P{3cm}|P{3cm}|P{3cm}|P{3cm}|}
        \hline
        \textbf{ID} & \textbf{Use-Case} & \textbf{SapBERT} & \textbf{KRISS BERT} & \textbf{BioBERT-snomed } & \textbf{CDE-Mapper (Llama3.1)} \\
        \hline
        I & hypercholesterolemia as a comorbid condition & secondary hypercholesterolemia  & hypercholesterolemia \faBook  & secondary hypercholesterolemia &   hypercholesterolemia \faBook \\
        \hline
        II & pmol/L & micromole per liter & micromole per liter  & picogram per milliliter & picomole per liter \faBook \\
        \hline
        III & N-terminal pro b-type natriuretic peptide in blood & plasma n-terminal pro b-type natriuretic peptide concentration measurement & n-terminal pro b-type natriuretic peptide mass concentration in serum & natriuretic peptide.b prohormone n-terminal [mass/volume] in blood by immunoassay & natriuretic peptide.b prohormone n-terminal|substance concentration|moment in time|blood, serum or plasma \faBullseye \\
        \hline
        IV & heart failure worsening & Exacerbation of heart failure \faBullseye & systolic heart failure & heart failure+transplant cardiology letter & Exacerbation of heart failure \faBullseye \\
        \hline
        V & diuretics dose expressed in an amount equivalent to Furosemide & furosemide | dose | drug doses \faBullseye & diuretic dose increasing & furosemide | dose | drug doses \faBullseye & furosemide | dose | drug doses \faBullseye \\
        \hline
         VI & o.i.d & o.d. unit	& 	o.d. unit & 1/day \faBullseye & once a day \faBullseye \\
        \hline
        VII & nonspecific t wave abnormality, improved in on ecg & nonspecific st-t abnormality on electrocardiogram	& 	nonspecific st-t abnormality on electrocardiogram & onspecific st-t abnormality on electrocardiogram & nonspecific st-t abnormality on electrocardiogram  \\
        \hline
         VIII & date of baseline visit & birth date	& date of visit & date of visit & date of first visit \faBullseye \\
        \hline
         IX & mmhg & millimeter	& hg & liter per minute &  millimeter mercury column \faBullseye \\
         \hline
        X & nyha class in half increments & new york heart association classification class \faBullseye	& heart failure with mid range ejection fraction & culaea inconstans &  new york heart association classification class \faBullseye \\
        \hline
         XI & heart rate measured in recumbent position & resting heart rate & pulse rate measured by palpation & heart rate unspecified time mean by pedometer & heart rate|recumbent body position \faBullseye \\
        \hline
         XII & mild to moderate asthma exacerbation & exacerbation of mild persistent asthma & moderate persistent asthma & exacerbation of mild persistent asthma & Acute asthma| mild to moderate \faBullseye \\
         \hline
         XIII & patient followed up 3 days after discharge & post hospital discharge followed up within 3 days \faBook & post hospital discharge followed up within 3 days \faBook & post hospital discharge followed up within 3 days \faBook & post hospital discharge followed up within 3 days \faBook \\
         \hline
         XIV & geniculate herpes zoster & herpes zoster oticus \faBullseye & herpes simplex keratitis & herpes zoster auricularis \faBullseye & Ramsay Hunt Syndrome Type II \faBullseye \\
         \hline
         XV & man & humanist & bisexual & male \faBook  & male \faBook \\
         \hline
         XVI & NYHA class: No limitation of physical activity & new york heart association classification - class i \faBullseye & new york heart association classification - class iv & new york heart association classification - class ii & new york heart association classification - class i \faBullseye \\
         \hline
         XVII & vena cava inferior max diameter on echocardiography & diameter|inferior vena cava|cardiac ultrasound \faBullseye & diameter|inferior vena cava|cardiac ultrasound \faBullseye & superior vena cava diameter by us & diameter | inferior vena cava|cardiac ultrasound  \faBullseye \\
         \hline
        \end{tabular}
     \begin{tablenotes}
        \footnotesize
        \item \faBook\ indicates the ground truth prediction.
        \item \faBullseye\ indicates predictions considered correct by clinicians.
    \end{tablenotes}
    \end{threeparttable}
    \caption{Model Predictions with Clinician and Ground Truth Justifications}
    \label{tab:model-predictions}
   
\end{table*}
\end{landscape}
\clearpage 

\bibliographystyle{unsrtnat}
\bibliography{references}  

\begin{thebibliography}{73}
\providecommand{\natexlab}[1]{#1}
\providecommand{\url}[1]{\texttt{#1}}
\expandafter\ifx\csname urlstyle\endcsname\relax
  \providecommand{\doi}[1]{doi: #1}\else
  \providecommand{\doi}{doi: \begingroup \urlstyle{rm}\Url}\fi

\bibitem[Kalankesh and Monaghesh(2024)]{kalankesh2024utilization}
Leila~R Kalankesh and Elham Monaghesh.
\newblock Utilization of ehrs for clinical trials: a systematic review.
\newblock \emph{BMC Medical Research Methodology}, 24\penalty0 (1):\penalty0 70, 2024.

\bibitem[Hutchings et~al.(2020)Hutchings, Loomes, Butow, and Boyle]{Hutchings2020A}
E.~Hutchings, Max~William Loomes, P.~Butow, and F.~Boyle.
\newblock A systematic literature review of researchers’ and healthcare professionals’ attitudes towards the secondary use and sharing of health administrative and clinical trial data.
\newblock \emph{Systematic Reviews}, 9, 2020.
\newblock \doi{10.1186/s13643-020-01485-5}.

\bibitem[Whitlock et~al.(2024)Whitlock, Leroy, Donovan, and Galgiani]{whitlock2024icd}
Abigail~E Whitlock, Gondy Leroy, Fariba~M Donovan, and John~N Galgiani.
\newblock Icd codes are insufficient to create datasets for machine learning: An evaluation using all of us data for coccidioidomycosis and myocardial infarction.
\newblock In \emph{2024 IEEE 12th International Conference on Healthcare Informatics (ICHI)}, pages 129--134. IEEE, 2024.

\bibitem[Hardy et~al.(2022)Hardy, Heyl, Tucker, Hopper, Marchã, Briggs, Yates, Day, Wheeler, Eve-Jones, and Gray]{Hardy2022DataConsistency}
Flavien Hardy, Johannes Heyl, Katie Tucker, Adrian Hopper, Maria~J Marchã, Tim W~R Briggs, Jeremy Yates, Jamie Day, Andrew Wheeler, Sue Eve-Jones, and William~K Gray.
\newblock Data consistency in the english hospital episodes statistics database.
\newblock \emph{BMJ Health Care Informatics}, 29\penalty0 (1):\penalty0 e100633, 2022.
\newblock \doi{10.1136/bmjhci-2022-100633}.

\bibitem[French and McInnes(2023)]{entity_linking_overview}
Evan French and Bridget~T McInnes.
\newblock An overview of biomedical entity linking throughout the years.
\newblock \emph{Journal of biomedical informatics}, 137:\penalty0 104252, 2023.

\bibitem[Shen et~al.(2021)Shen, Li, Liu, Han, Wang, and Yuan]{Shen2021Entity}
Wei Shen, Yuhan Li, Yinan Liu, Jiawei Han, Jianyong Wang, and Xiaojie Yuan.
\newblock Entity linking meets deep learning: Techniques and solutions.
\newblock \emph{IEEE Transactions on Knowledge and Data Engineering}, 35:\penalty0 2556--2578, 2021.
\newblock \doi{10.1109/TKDE.2021.3117715}.

\bibitem[Guven and Lamurias(2023)]{Guven2023Multilingual}
Zekeriya~Anil Guven and Andre Lamurias.
\newblock Multilingual bi‐encoder models for biomedical entity linking.
\newblock \emph{Expert Systems}, 40, 2023.
\newblock \doi{10.1111/exsy.13388}.

\bibitem[Cutrona et~al.(2020)Cutrona, Bianchi, Jim{\'e}nez-Ruiz, and Palmonari]{tabulardataEL}
Vincenzo Cutrona, Federico Bianchi, Ernesto Jim{\'e}nez-Ruiz, and Matteo Palmonari.
\newblock Tough tables: Carefully evaluating entity linking for tabular data.
\newblock In \emph{International Semantic Web Conference}, pages 328--343. Springer, 2020.

\bibitem[Kulyabin et~al.(2024)Kulyabin, Sokolov, Galaida, Maier, and Arias-Vergara]{SNOBERT}
Mikhail Kulyabin, Gleb Sokolov, Aleksandr Galaida, Andreas Maier, and Tomas Arias-Vergara.
\newblock Snobert: A benchmark for clinical notes entity linking in the snomed ct clinical terminology.
\newblock \emph{arXiv preprint arXiv:2405.16115}, 2024.

\bibitem[Houssein et~al.(2021)Houssein, Mohamed, and Ali]{ML_Bio_NLP}
E.~H. Houssein, Rehab~E. Mohamed, and Abdelmgeid~A. Ali.
\newblock Machine learning techniques for biomedical natural language processing: A comprehensive review.
\newblock \emph{IEEE Access}, PP:\penalty0 1--1, 2021.
\newblock \doi{10.1109/ACCESS.2021.3119621}.

\bibitem[Loureiro and Jorge(2020)]{Loureiro2020MedLinker}
Daniel Loureiro and A.~Jorge.
\newblock Medlinker: Medical entity linking with neural representations and dictionary matching.
\newblock \emph{Advances in Information Retrieval}, 12036:\penalty0 230 -- 237, 2020.
\newblock \doi{10.1007/978-3-030-45442-5_29}.

\bibitem[Perera et~al.(2020)Perera, Dehmer, and Emmert-Streib]{Perera2020Named}
Nadeesha Perera, M.~Dehmer, and F.~Emmert-Streib.
\newblock Named entity recognition and relation detection for biomedical information extraction.
\newblock \emph{Frontiers in Cell and Developmental Biology}, 8, 2020.
\newblock \doi{10.3389/fcell.2020.00673}.

\bibitem[Abdullahi et~al.(2024)Abdullahi, Mercurio, Singh, and Eickhoff]{retrieval_based_diagnostic_support}
Tassallah Abdullahi, Laura Mercurio, Ritambhara Singh, and Carsten Eickhoff.
\newblock Retrieval-based diagnostic decision support: Mixed methods study.
\newblock \emph{JMIR Medical Informatics}, 12:\penalty0 e50209, 2024.

\bibitem[Soldaini and Goharian(2016)]{quickumls}
Luca Soldaini and Nazli Goharian.
\newblock Quickumls: a fast, unsupervised approach for medical concept extraction.
\newblock In \emph{MedIR workshop, sigir}, pages 1--4, 2016.

\bibitem[Lhncbc()]{MetaMapLite}
Lhncbc.
\newblock Lhncbc/metamaplite: A near real-time named-entity recognizer.
\newblock URL \url{https://github.com/lhncbc/metamaplite}.

\bibitem[Savova et~al.(2010)Savova, Masanz, Ogren, Zheng, Sohn, Kipper-Schuler, and Chute]{CTAKES}
Guergana~K Savova, James~J Masanz, Philip~V Ogren, Jiaping Zheng, Sunghwan Sohn, Karin~C Kipper-Schuler, and Christopher~G Chute.
\newblock Mayo clinical text analysis and knowledge extraction system (ctakes): Architecture, component evaluation and applications.
\newblock \emph{Journal of the American Medical Informatics Association}, 17\penalty0 (5):\penalty0 507–513, Sep 2010.
\newblock \doi{10.1136/jamia.2009.001560}.

\bibitem[Zhao et~al.(2023)Zhao, Zhou, Li, Tang, Wang, Hou, Min, Zhang, Zhang, Dong, et~al.]{zhao2023survey}
Wayne~Xin Zhao, Kun Zhou, Junyi Li, Tianyi Tang, Xiaolei Wang, Yupeng Hou, Yingqian Min, Beichen Zhang, Junjie Zhang, Zican Dong, et~al.
\newblock A survey of large language models.
\newblock \emph{arXiv preprint arXiv:2303.18223}, 2023.

\bibitem[Naveed et~al.(2023)Naveed, Khan, Qiu, Saqib, Anwar, Usman, Akhtar, Barnes, and Mian]{naveed2023comprehensive}
Humza Naveed, Asad~Ullah Khan, Shi Qiu, Muhammad Saqib, Saeed Anwar, Muhammad Usman, Naveed Akhtar, Nick Barnes, and Ajmal Mian.
\newblock A comprehensive overview of large language models.
\newblock \emph{arXiv preprint arXiv:2307.06435}, 2023.

\bibitem[Thirunavukarasu et~al.(2023)Thirunavukarasu, Ting, Elangovan, Gutierrez, Tan, and Ting]{thirunavukarasu2023large}
Arun~James Thirunavukarasu, Darren Shu~Jeng Ting, Kabilan Elangovan, Laura Gutierrez, Ting~Fang Tan, and Daniel Shu~Wei Ting.
\newblock Large language models in medicine.
\newblock \emph{Nature medicine}, 29\penalty0 (8):\penalty0 1930--1940, 2023.

\bibitem[Chang et~al.(2024)Chang, Wang, Wang, Wu, Yang, Zhu, Chen, Yi, Wang, Wang, et~al.]{chang2024survey}
Yupeng Chang, Xu~Wang, Jindong Wang, Yuan Wu, Linyi Yang, Kaijie Zhu, Hao Chen, Xiaoyuan Yi, Cunxiang Wang, Yidong Wang, et~al.
\newblock A survey on evaluation of large language models.
\newblock \emph{ACM Transactions on Intelligent Systems and Technology}, 15\penalty0 (3):\penalty0 1--45, 2024.

\bibitem[Lewis et~al.(2020)Lewis, Perez, Piktus, Petroni, Karpukhin, Goyal, Kuttler, Lewis, tau Yih, Rocktäschel, Riedel, and Kiela]{Lewis2020Retrieval-Augmented}
Patrick Lewis, Ethan Perez, Aleksandara Piktus, Fabio Petroni, Vladimir Karpukhin, Naman Goyal, Heinrich Kuttler, M.~Lewis, Wen tau Yih, Tim Rocktäschel, Sebastian Riedel, and Douwe Kiela.
\newblock Retrieval-augmented generation for knowledge-intensive nlp tasks.
\newblock \emph{ArXiv}, abs/2005.11401, 2020.

\bibitem[Thapa and Adhikari(2023)]{thapa2023chatgpt}
Surendrabikram Thapa and Surabhi Adhikari.
\newblock Chatgpt, bard, and large language models for biomedical research: opportunities and pitfalls.
\newblock \emph{Annals of biomedical engineering}, 51\penalty0 (12):\penalty0 2647--2651, 2023.

\bibitem[Caufield et~al.(2024)Caufield, Hegde, Emonet, Harris, Joachimiak, Matentzoglu, Kim, Moxon, Reese, Haendel, Robinson, and Mungall]{SPIRES}
J~Harry Caufield, Harshad Hegde, Vincent Emonet, Nomi~L Harris, Marcin~P Joachimiak, Nicolas Matentzoglu, HyeongSik Kim, Sierra Moxon, Justin~T Reese, Melissa~A Haendel, Peter~N Robinson, and Christopher~J Mungall.
\newblock Structured prompt interrogation and recursive extraction of semantics (spires): a method for populating knowledge bases using zero-shot learning.
\newblock \emph{Bioinformatics}, 40\penalty0 (3), February 2024.
\newblock ISSN 1367-4811.
\newblock \doi{10.1093/bioinformatics/btae104}.
\newblock URL \url{http://dx.doi.org/10.1093/bioinformatics/btae104}.

\bibitem[Zhang et~al.(2023)Zhang, Li, Cui, Cai, Liu, Fu, Huang, Zhao, Zhang, Chen, Wang, Luu, Bi, Shi, and Shi]{Zhang2023Siren's}
Yue Zhang, Yafu Li, Leyang Cui, Deng Cai, Lemao Liu, Tingchen Fu, Xinting Huang, Enbo Zhao, Yu~Zhang, Yulong Chen, Longyue Wang, A.~Luu, Wei Bi, Freda Shi, and Shuming Shi.
\newblock Siren's song in the ai ocean: A survey on hallucination in large language models.
\newblock \emph{ArXiv}, abs/2309.01219, 2023.
\newblock \doi{10.48550/arXiv.2309.01219}.

\bibitem[Gao et~al.(2024)Gao, Xiong, Gao, Jia, Pan, Bi, Dai, Sun, Wang, and Wang]{gao2024retrievalaugmentedgenerationlargelanguage}
Yunfan Gao, Yun Xiong, Xinyu Gao, Kangxiang Jia, Jinliu Pan, Yuxi Bi, Yi~Dai, Jiawei Sun, Meng Wang, and Haofen Wang.
\newblock Retrieval-augmented generation for large language models: A survey.
\newblock \emph{arXiv preprint arXiv:2312.10997}, 2024.
\newblock URL \url{https://arxiv.org/abs/2312.10997}.

\bibitem[Liu et~al.(2021{\natexlab{a}})Liu, Shareghi, Meng, Basaldella, and Collier]{liu2021selfalignment}
Fangyu Liu, Ehsan Shareghi, Zaiqiao Meng, Marco Basaldella, and Nigel Collier.
\newblock Self-alignment pretraining for biomedical entity representations.
\newblock In \emph{Proceedings of the 2021 Conference of the North American Chapter of the Association for Computational Linguistics: Human Language Technologies}, 2021{\natexlab{a}}.

\bibitem[Xie et~al.(2024)Xie, Lu, Ho, Nahab, Hu, and Yang]{promptlink_MCN}
Yuzhang Xie, Jiaying Lu, Joyce Ho, Fadi Nahab, Xiao Hu, and Carl Yang.
\newblock Promptlink: Leveraging large language models for cross-source biomedical concept linking.
\newblock \emph{arXiv preprint arXiv:2405.07500}, 2024.

\bibitem[Zhang et~al.(2021)Zhang, Cheng, Vashishth, Wong, Xiao, Liu, Naumann, Gao, and Poon]{kriss_bert}
Sheng Zhang, Hao Cheng, Shikhar Vashishth, Cliff Wong, Jinfeng Xiao, Xiaodong Liu, Tristan Naumann, Jianfeng Gao, and Hoifung Poon.
\newblock Knowledge-rich self-supervision for biomedical entity linking.
\newblock \emph{arXiv preprint arXiv:2112.07887}, 2021.

\bibitem[Sheehan et~al.(2016)Sheehan, Hirschfeld, Foster, Ghitza, Goetz, Karpinski, Lang, Moser, Odenkirchen, Reeves, et~al.]{sheehan2016improving}
Jerry Sheehan, Steven Hirschfeld, Erin Foster, Udi Ghitza, Kerry Goetz, Joanna Karpinski, Lisa Lang, Richard~P Moser, Joanne Odenkirchen, Dianne Reeves, et~al.
\newblock Improving the value of clinical research through the use of common data elements.
\newblock \emph{Clinical Trials}, 13\penalty0 (6):\penalty0 671--676, 2016.

\bibitem[Le~Sueur et~al.(2020)Le~Sueur, Bruce, Geifman, and Consortium]{le2020challenges}
Helen Le~Sueur, Ian~N Bruce, Nophar Geifman, and Masterplans Consortium.
\newblock The challenges in data integration--heterogeneity and complexity in clinical trials and patient registries of systemic lupus erythematosus.
\newblock \emph{BMC medical research methodology}, 20:\penalty0 1--5, 2020.

\bibitem[Kim et~al.(2020)Kim, Park, Lee, and Kim]{compositeCDEs}
Hye~Hyeon Kim, Yu~Rang Park, Suehyun Lee, and Ju~Han Kim.
\newblock Composite cde: modeling composite relationships between common data elements for representing complex clinical data.
\newblock \emph{BMC Medical Informatics and Decision Making}, 20:\penalty0 1--18, 2020.

\bibitem[Platt et~al.(2018)Platt, Brown, Robb, McClellan, Ball, Nguyen, and Sherman]{platt2018fda}
Richard Platt, Jeffrey~S Brown, Melissa Robb, Mark McClellan, Robert Ball, Michael~D Nguyen, and Rachel~E Sherman.
\newblock The fda sentinel initiative—an evolving national resource.
\newblock \emph{N Engl J Med}, 379\penalty0 (22):\penalty0 2091--2093, 2018.

\bibitem[Klann et~al.(2019)Klann, Joss, Embree, and Murphy]{klann2019data}
Jeffrey~G Klann, Matthew~AH Joss, Kevin Embree, and Shawn~N Murphy.
\newblock Data model harmonization for the all of us research program: Transforming i2b2 data into the omop common data model.
\newblock \emph{PloS one}, 14\penalty0 (2):\penalty0 e0212463, 2019.

\bibitem[Stang et~al.(2010)Stang, Ryan, Racoosin, Overhage, Hartzema, Reich, Welebob, Scarnecchia, and Woodcock]{OMOP_advancement}
Paul~E Stang, Patrick~B Ryan, Judith~A Racoosin, J~Marc Overhage, Abraham~G Hartzema, Christian Reich, Emily Welebob, Thomas Scarnecchia, and Janet Woodcock.
\newblock Advancing the science for active surveillance: rationale and design for the observational medical outcomes partnership.
\newblock \emph{Annals of internal medicine}, 153\penalty0 (9):\penalty0 600--606, 2010.

\bibitem[Lehne et~al.(2019)Lehne, Sass, Essenwanger, Schepers, and Thun]{interoperability_in_digital_medicine}
Moritz Lehne, Julian Sass, Andrea Essenwanger, Josef Schepers, and Sylvia Thun.
\newblock Why digital medicine depends on interoperability.
\newblock \emph{NPJ digital medicine}, 2\penalty0 (1):\penalty0 79, 2019.

\bibitem[Reps et~al.(2018)Reps, Schuemie, Suchard, Ryan, and Rijnbeek]{design_implemented_standardized_framework_patient_level_prediction}
Jenna~M Reps, Martijn~J Schuemie, Marc~A Suchard, Patrick~B Ryan, and Peter~R Rijnbeek.
\newblock Design and implementation of a standardized framework to generate and evaluate patient-level prediction models using observational healthcare data.
\newblock \emph{Journal of the American Medical Informatics Association}, 25\penalty0 (8):\penalty0 969--975, 2018.

\bibitem[Papez et~al.(2021)Papez, Moinat, Payralbe, Asselbergs, Lumbers, Hemingway, Dobson, and Denaxas]{omop_EHR_phenotyping}
Vaclav Papez, Maxim Moinat, Stefan Payralbe, Folkert~W Asselbergs, R~Thomas Lumbers, Harry Hemingway, Richard Dobson, and Spiros Denaxas.
\newblock Transforming and evaluating electronic health record disease phenotyping algorithms using the omop common data model: a case study in heart failure.
\newblock \emph{JAMIA open}, 4\penalty0 (3):\penalty0 ooab001, 2021.

\bibitem[Ahmadi et~al.(2024)Ahmadi, Zoch, Guengoeze, Facchinello, Mondorf, Stratmann, Musleh, Erasmus, Tchertov, Gebler, et~al.]{omop_rare_diseases}
Najia Ahmadi, Michele Zoch, Oya Guengoeze, Carlo Facchinello, Antonia Mondorf, Katharina Stratmann, Khader Musleh, Hans-Peter Erasmus, Jana Tchertov, Richard Gebler, et~al.
\newblock How to customize common data models for rare diseases: an omop-based implementation and lessons learned.
\newblock \emph{Orphanet Journal of Rare Diseases}, 19\penalty0 (1):\penalty0 298, 2024.

\bibitem[Biedermann et~al.(2021)Biedermann, Ong, Davydov, Orlova, Solovyev, Sun, Wetherill, Brand, and Didden]{omop_standardizing_pulmonary_hypertension}
Patricia Biedermann, Rose Ong, Alexander Davydov, Alexandra Orlova, Philip Solovyev, Hong Sun, Graham Wetherill, Monika Brand, and Eva-Maria Didden.
\newblock Standardizing registry data to the omop common data model: experience from three pulmonary hypertension databases.
\newblock \emph{BMC medical research methodology}, 21:\penalty0 1--16, 2021.

\bibitem[Izacard et~al.(2022)Izacard, Lewis, Lomeli, Hosseini, Petroni, Schick, Dwivedi-Yu, Joulin, Riedel, and Grave]{few_shot_RAG}
Gautier Izacard, Patrick Lewis, Maria Lomeli, Lucas Hosseini, Fabio Petroni, Timo Schick, Jane Dwivedi-Yu, Armand Joulin, Sebastian Riedel, and Edouard Grave.
\newblock Few-shot learning with retrieval augmented language models.
\newblock \emph{arXiv preprint arXiv:2208.03299}, 2\penalty0 (3), 2022.

\bibitem[Rubin et~al.(2021)Rubin, Herzig, and Berant]{rubin2021learning}
Ohad Rubin, Jonathan Herzig, and Jonathan Berant.
\newblock Learning to retrieve prompts for in-context learning.
\newblock \emph{arXiv preprint arXiv:2112.08633}, 2021.

\bibitem[Dong et~al.(2022{\natexlab{a}})Dong, Li, Dai, Zheng, Wu, Chang, Sun, Xu, and Sui]{survey_in_context_learning}
Qingxiu Dong, Lei Li, Damai Dai, Ce~Zheng, Zhiyong Wu, Baobao Chang, Xu~Sun, Jingjing Xu, and Zhifang Sui.
\newblock A survey on in-context learning.
\newblock \emph{arXiv preprint arXiv:2301.00234}, 2022{\natexlab{a}}.

\bibitem[Dong et~al.(2022{\natexlab{b}})Dong, Li, Dai, Zheng, Ma, Li, Xia, Xu, Wu, Liu, et~al.]{ICL_survey}
Qingxiu Dong, Lei Li, Damai Dai, Ce~Zheng, Jingyuan Ma, Rui Li, Heming Xia, Jingjing Xu, Zhiyong Wu, Tianyu Liu, et~al.
\newblock A survey on in-context learning.
\newblock \emph{arXiv preprint arXiv:2301.00234}, 2022{\natexlab{b}}.

\bibitem[Liu et~al.(2021{\natexlab{b}})Liu, Shen, Zhang, Dolan, Carin, and Chen]{incontext_examples_gpt3}
Jiachang Liu, Dinghan Shen, Yizhe Zhang, Bill Dolan, Lawrence Carin, and Weizhu Chen.
\newblock What makes good in-context examples for gpt-$3 $?
\newblock \emph{arXiv preprint arXiv:2101.06804}, 2021{\natexlab{b}}.

\bibitem[Shi et~al.(2022)Shi, Michael, Gururangan, and Zettlemoyer]{knn_prompting}
Weijia Shi, Julian Michael, Suchin Gururangan, and Luke Zettlemoyer.
\newblock knn-prompt: Nearest neighbor zero-shot inference.
\newblock \emph{arXiv preprint arXiv:2205.13792}, 2022.

\bibitem[Chen et~al.(2024)Chen, Lin, Han, and Sun]{benchamrking_RAG}
Jiawei Chen, Hongyu Lin, Xianpei Han, and Le~Sun.
\newblock Benchmarking large language models in retrieval-augmented generation (2023).
\newblock \emph{arXiv preprint arXiv:2309.01431}, 2024.

\bibitem[Wolf et~al.(2019)Wolf, Debut, Sanh, Chaumond, Delangue, Moi, Cistac, Rault, Louf, Funtowicz, et~al.]{wolf2019huggingface}
Thomas Wolf, Lysandre Debut, Victor Sanh, Julien Chaumond, Clement Delangue, Anthony Moi, Pierric Cistac, Tim Rault, R{\'e}mi Louf, Morgan Funtowicz, et~al.
\newblock Huggingface's transformers: State-of-the-art natural language processing.
\newblock \emph{arXiv preprint arXiv:1910.03771}, 2019.

\bibitem[Chandrasekaran and Mago(2021)]{semantic_Similarity_evoluation}
Dhivya Chandrasekaran and Vijay Mago.
\newblock Evolution of semantic similarity—a survey.
\newblock \emph{ACM Computing Surveys (CSUR)}, 54\penalty0 (2):\penalty0 1--37, 2021.

\bibitem[Formal et~al.(2021)Formal, Lassance, Piwowarski, and Clinchant]{spladev2}
Thibault Formal, Carlos Lassance, Benjamin Piwowarski, and St{\'{e}}phane Clinchant.
\newblock {SPLADE} v2: Sparse lexical and expansion model for information retrieval.
\newblock \emph{CoRR}, abs/2109.10086, 2021.
\newblock URL \url{https://arxiv.org/abs/2109.10086}.

\bibitem[Song et~al.(2021)Song, Li, Liu, and Zeng]{DP_bioNER}
Bosheng Song, Fen Li, Yuansheng Liu, and Xiangxiang Zeng.
\newblock Deep learning methods for biomedical named entity recognition: a survey and qualitative comparison.
\newblock \emph{Briefings in bioinformatics}, 2021.
\newblock \doi{10.1093/bib/bbab282}.

\bibitem[Sung et~al.(2020{\natexlab{a}})Sung, Jeon, Lee, and Kang]{bib3}
Mujeen Sung, Hwisang Jeon, Jinhyuk Lee, and Jaewoo Kang.
\newblock Biomedical entity representations with synonym marginalization.
\newblock In Dan Jurafsky, Joyce Chai, Natalie Schluter, and Joel Tetreault, editors, \emph{Proceedings of the 58th Annual Meeting of the Association for Computational Linguistics}, pages 3641--3650, Online, jul 2020{\natexlab{a}}. Association for Computational Linguistics.
\newblock \doi{10.18653/v1/2020.acl-main.335}.
\newblock URL \url{https://aclanthology.org/2020.acl-main.335}.

\bibitem[Yan et~al.(2021)Yan, Zhang, Liu, Zhao, Shi, and Liu]{bib2}
Cheng Yan, Yuanzhe Zhang, Kang Liu, Jun Zhao, Yafei Shi, and Shengping Liu.
\newblock Biomedical concept normalization by leveraging hypernyms.
\newblock In \emph{Proceedings of the 2021 conference on empirical methods in natural language processing}, pages 3512--3517, 2021.

\bibitem[Zheng et~al.(2014)Zheng, Howsmon, Zhang, Hahn, McGuinness, Hendler, and Ji]{Zheng2014Entity}
Jinguang Zheng, D.~Howsmon, Boliang Zhang, J.~Hahn, D.~McGuinness, J.~Hendler, and Heng Ji.
\newblock Entity linking for biomedical literature.
\newblock \emph{BMC Medical Informatics and Decision Making}, 15:\penalty0 S4 -- S4, 2014.
\newblock \doi{10.1186/1472-6947-15-S1-S4}.

\bibitem[Lee et~al.(2020)Lee, Yoon, Kim, Kim, Kim, So, and Kang]{lee2020biobert}
Jinhyuk Lee, Wonjin Yoon, Sungdong Kim, Donghyeon Kim, Sunkyu Kim, Chan~Ho So, and Jaewoo Kang.
\newblock Biobert: a pre-trained biomedical language representation model for biomedical text mining.
\newblock \emph{Bioinformatics}, 36\penalty0 (4):\penalty0 1234--1240, 2020.

\bibitem[Zhu et~al.(2022)Zhu, Qin, Chen, Hu, and Xiang]{enhancing_entity_rep_prompt_learning_bioent}
Tiantian Zhu, Yang Qin, Qingcai Chen, Baotian Hu, and Yang Xiang.
\newblock Enhancing entity representations with prompt learning for biomedical entity linking.
\newblock In \emph{Proceedings of the 31st International Joint Conference on Artificial Intelligence (IJCAI-22)}, pages 4036--4042, 2022.
\newblock \doi{10.24963/ijcai.2022/560}.

\bibitem[Zhu et~al.(2023)Zhu, Qin, Feng, Chen, Hu, and Xiang]{biopro}
Tiantian Zhu, Yang Qin, Ming Feng, Qingcai Chen, Baotian Hu, and Yang Xiang.
\newblock Biopro: Context-infused prompt learning for biomedical entity linking.
\newblock \emph{IEEE/ACM Transactions on Audio, Speech, and Language Processing}, 32:\penalty0 374--385, 2023.

\bibitem[Wang et~al.(2020)Wang, Wang, Bai, Liu, Liu, Zhang, Jiang, Xu, Wang, and Zhou]{ehr2vec}
Li~Wang, Qinghua Wang, Heming Bai, Cong Liu, Wei Liu, Yuanpeng Zhang, Lei Jiang, Huji Xu, Kai Wang, and Yunyun Zhou.
\newblock Ehr2vec: representation learning of medical concepts from temporal patterns of clinical notes based on self-attention mechanism.
\newblock \emph{Frontiers in Genetics}, 11:\penalty0 630, 2020.

\bibitem[Achiam et~al.(2023)Achiam, Adler, Agarwal, Ahmad, Akkaya, Aleman, Almeida, Altenschmidt, Altman, Anadkat, et~al.]{achiam2023gpt}
Josh Achiam, Steven Adler, Sandhini Agarwal, Lama Ahmad, Ilge Akkaya, Florencia~Leoni Aleman, Diogo Almeida, Janko Altenschmidt, Sam Altman, Shyamal Anadkat, et~al.
\newblock Gpt-4 technical report.
\newblock \emph{arXiv preprint arXiv:2303.08774}, 2023.

\bibitem[Touvron et~al.(2023)Touvron, Lavril, Izacard, Martinet, Lachaux, Lacroix, Rozi{\`e}re, Goyal, Hambro, Azhar, et~al.]{touvron2023llama}
Hugo Touvron, Thibaut Lavril, Gautier Izacard, Xavier Martinet, Marie-Anne Lachaux, Timoth{\'e}e Lacroix, Baptiste Rozi{\`e}re, Naman Goyal, Eric Hambro, Faisal Azhar, et~al.
\newblock Llama: Open and efficient foundation language models.
\newblock \emph{arXiv preprint arXiv:2302.13971}, 2023.

\bibitem[Soman et~al.(2023)Soman, Rose, Morris, E~Akbas, Smith, Peetoom, Villouta-Reyes, Cerono, Shi, Rizk-Jackson, et~al.]{biokg_llm}
Karthik Soman, Peter~W Rose, John~H Morris, Rabia E~Akbas, Brett Smith, Braian Peetoom, Catalina Villouta-Reyes, Gabriel Cerono, Yongmei Shi, Angela Rizk-Jackson, et~al.
\newblock Biomedical knowledge graph-enhanced prompt generation for large language models.
\newblock \emph{arXiv e-prints}, pages arXiv--2311, 2023.

\bibitem[Fernandez et~al.(2023)Fernandez, Elmore, Franklin, Krishnan, and Tan]{llm_data_management}
Raul~Castro Fernandez, Aaron~J Elmore, Michael~J Franklin, Sanjay Krishnan, and Chenhao Tan.
\newblock How large language models will disrupt data management.
\newblock \emph{Proceedings of the VLDB Endowment}, 16\penalty0 (11):\penalty0 3302--3309, 2023.

\bibitem[Peeters et~al.(2023)Peeters, Steiner, and Bizer]{entity_match_llm}
Ralph Peeters, Aaron Steiner, and Christian Bizer.
\newblock Entity matching using large language models.
\newblock \emph{arXiv preprint arXiv:2310.11244}, 2023.

\bibitem[Wang et~al.(2023)Wang, Gao, and Xu]{incontext_learning_BioNEL}
Qinyong Wang, Zhenxiang Gao, and Rong Xu.
\newblock Exploring the in-context learning ability of large language model for biomedical concept linking.
\newblock \emph{ArXiv}, abs/2307.01137, 2023.
\newblock \doi{10.48550/arXiv.2307.01137}.

\bibitem[Liu et~al.(2023)Liu, Chabot, Troncy, Huynh, Labb{\'e}, and Monnin]{liu2023tabular}
Jixiong Liu, Yoan Chabot, Rapha{\"e}l Troncy, Viet-Phi Huynh, Thomas Labb{\'e}, and Pierre Monnin.
\newblock From tabular data to knowledge graphs: A survey of semantic table interpretation tasks and methods.
\newblock \emph{Journal of Web Semantics}, 76:\penalty0 100761, 2023.

\bibitem[Li et~al.(2016)Li, Sun, Johnson, Sciaky, Wei, Leaman, Davis, Mattingly, Wiegers, and Lu]{Li2016BioCreative}
Jiao Li, Yueping Sun, Robin~J. Johnson, D.~Sciaky, Chih-Hsuan Wei, Robert Leaman, A.~P. Davis, C.~Mattingly, Thomas~C. Wiegers, and Zhiyong Lu.
\newblock Biocreative v cdr task corpus: a resource for chemical disease relation extraction.
\newblock \emph{Database: The Journal of Biological Databases and Curation}, 2016, 2016.
\newblock \doi{10.1093/database/baw068}.

\bibitem[Do{\u g}an et~al.(2014)Do{\u g}an, Leaman, and Lu]{NCBI}
Rezarta~Islamaj Do{\u g}an, Robert Leaman, and Zhiyong Lu.
\newblock {NCBI} disease corpus: A resource for disease name recognition and concept normalization.
\newblock \emph{J. Biomed. Inform.}, 47:\penalty0 1--10, February 2014.

\bibitem[Maeder(2018)]{maeder2018time_chf}
M.~T. Maeder.
\newblock \emph{The Trial of Intensified Medical Therapy in Elderly Patients with Congestive Heart Failure (TIME-CHF): Novel Insights into Hot Topics in Heart Failure}.
\newblock PhD thesis, Maastricht University, 2018.

\bibitem[Handoko and van~de Bovenkamp(2020)]{CardioMEMS}
M.~L. Handoko and A.~A. van~de Bovenkamp.
\newblock Cardiomems: the next revolution in heart failure management?
\newblock \emph{Netherlands Heart Journal}, 28\penalty0 (1):\penalty0 14--15, Jan 2020.
\newblock ISSN 1568-5888.
\newblock \doi{10.1007/s12471-019-01356-2}.
\newblock M.L. Handoko participates in the MONITOR-HF trial. A.A. van de Bovenkamp declares no competing interests.

\bibitem[Sung et~al.(2020{\natexlab{b}})Sung, Jeon, Lee, and Kang]{sung2020biomedical}
Mujeen Sung, Hwisang Jeon, Jinhyuk Lee, and Jaewoo Kang.
\newblock Biomedical entity representations with synonym marginalization.
\newblock \emph{arXiv preprint arXiv:2005.00239}, 2020{\natexlab{b}}.

\bibitem[Lab()]{OHDSIananke}
Panacea Lab.
\newblock {OHDSIananke}: {OHDSI} ananke - a tool for mapping between {OHDSI} concept identifiers to unified medical language system ({UMLS}) identifiers.

\bibitem[Kojima et~al.(2022)Kojima, Gu, Reid, Matsuo, and Iwasawa]{llmaszeroshot}
Takeshi Kojima, Shixiang~Shane Gu, Machel Reid, Yutaka Matsuo, and Yusuke Iwasawa.
\newblock Large language models are zero-shot reasoners.
\newblock \emph{Advances in neural information processing systems}, 35:\penalty0 22199--22213, 2022.

\bibitem[Liu et~al.(2024)Liu, Lin, Hewitt, Paranjape, Bevilacqua, Petroni, and Liang]{liu2024lost}
Nelson~F Liu, Kevin Lin, John Hewitt, Ashwin Paranjape, Michele Bevilacqua, Fabio Petroni, and Percy Liang.
\newblock Lost in the middle: How language models use long contexts.
\newblock \emph{Transactions of the Association for Computational Linguistics}, 12:\penalty0 157--173, 2024.

\bibitem[J{\"a}rvelin and Kek{\"a}l{\"a}inen(2002)]{ncgd_ir}
Kalervo J{\"a}rvelin and Jaana Kek{\"a}l{\"a}inen.
\newblock Cumulated gain-based evaluation of ir techniques.
\newblock \emph{ACM Transactions on Information Systems (TOIS)}, 20\penalty0 (4):\penalty0 422--446, 2002.

\end{thebibliography}
\appendix
\section{Alogrithm}
\label{app:algorithm_section}
\begin{algorithm}

\label{alg:normalization}
\begin{algorithmic}[1]
\Require Dictionary of source terms $D$, linking rules template $M$
\Ensure All sub-terms in $D$ are mapped to standard terminologies according to $M$

\State Initialize list $N \gets [\, ]$ \Comment{List to store linked concepts for each query}
\For{each query $q \in D$}
    \State $q \gets \text{ValidateInput}(q)$ \Comment{Ensure $q$ conforms to expected format}

    \State $result \gets \text{DecomposeQuery}(q)$ \Comment{Decompose query into components $q_i$}
    \If{$result \neq \emptyset$}
        \State Initialize \textbf{candidate\_result} $\gets$ \{\} \Comment{Dictionary to store candidates for $q$}
        \For{each component $q_i$ in $result$}
            \State $kr\_match \gets \text{CheckInKR}(q_i)$ \Comment{Check if candidates exists in Knowledge Reservoir}
            \If{$kr\_match \neq \emptyset$}
                \State \textbf{candidate\_result[$q_i$]} $\gets$ $kr\_match$ \Comment{Use existing candidates}
            \Else
                \State $candidates \gets \text{MergeRetrievers}(q_i)$ \Comment{Retrieve potential candidates}
                \State $candidates \gets \text{KG\_Filter}(candidates, q_i, \text{rules})$ \Comment{Filter candidates based on expert rules and semantic relevance}
                \If{$candidates \neq \emptyset$}
                    \State $match \gets \text{FindExactMatch}(candidates, q_i)$
                    \If{$match \neq \emptyset$}
                        \State \textbf{candidate\_result[$q_i$]} $\gets$ $match$ \Comment{Add to result for $q$}
                    \Else
                        \State $scores \gets \text{Multistep\_LLMScore}(candidates)$ \Comment{Re-rank candidates based on LLM scoring}
                        \State $selected\_candidate \gets \text{SelectTopCandidate}(scores)$ \Comment{Select top candidate}
                        \State \textbf{candidate\_result[$q_i$]} $\gets$ $selected\_candidate$
                    \EndIf
                \Else
                    \State \textbf{candidate\_result[$q_i$]} $\gets$ ``NA" \Comment{No suitable candidates found}
                \EndIf
                \State $\text{AddInKR}(q_i, \textbf{candidate\_result[$q_i$]})$ \Comment{Add new candidate to Knowledge Reservoir}
            \EndIf
        \EndFor
        \State $N \gets N \cup \{\textbf{candidate\_result}\}$ \Comment{Add candidate result for $q$ to $N$}
    \Else
        \State $N \gets N \cup \{\text{``NA"}\}$ \Comment{Add ``NA" if decomposition fails}
    \EndIf
\EndFor

\Return $N$
\end{algorithmic}
\caption{Clinical Terms Linking Process}
\end{algorithm}






\end{document}